\documentclass[journal,final,onecolumn,12pt,twoside]{IEEEtranTCOM}
\normalsize
\usepackage{cite}
\usepackage{graphicx}
\usepackage{subfigure}
\usepackage{amsmath}
\usepackage{amssymb}
\usepackage{bm}
\usepackage{epsfig}
\usepackage{times}
\usepackage{color}
\usepackage{array}
\usepackage{multirow}
\usepackage{rotating}
\usepackage{extarrows}
\usepackage{algpseudocode}
\usepackage[ruled, vlined, linesnumbered]{algorithm2e}
\usepackage{tabularx}
\usepackage{epstopdf}
\usepackage{amsfonts}
\usepackage{lipsum}
\usepackage{mathtools}
\usepackage{cuted}
\usepackage{setspace}
\usepackage{multicol}

\usepackage{kantlipsum}

\usepackage{float}
\date{}
\def\IEEEQEDclosed{\mbox{\rule[0pt]{1.3ex}{1.3ex}}} 
\def\IEEEQEDopen{{\setlength{\fboxsep}{0pt}\setlength{\fboxrule}{0.2pt}\fbox{\rule[0pt]{0pt}{1.3ex}\rule[0pt]{1.3ex}{0pt}}}}

\def\@begintheorem#1#2{\par\bgroup{\scshape #1\ #2. }\it\ignorespaces}
\def\@opargbegintheorem#1#2#3{\par\bgroup%
   {\scshape #1\ #2\ ({\upshape #3}). }\it\ignorespaces}
\def\@endtheorem{\egroup}

\newtheorem{pavikl}{\textbf{Lemma}}

\newtheorem{pavikp}{\textbf{Proposition}}

\newcommand{\argmin}{\operatornamewithlimits{argmin}}

\renewcommand{\arraystretch}{1.5}

\makeatletter

\newcommand{\Rmnum}[1]{\expandafter\@slowromancap\romannumeral #1@}
\makeatother

\begin{document}
\title{Multiple D2D Multicasts in Underlay Cellular Networks}
\author{\IEEEauthorblockN{Ajay Bhardwaj and Samar Agnihotri}%

\IEEEauthorblockA{School of Computing and Electrical Engineering, IIT Mandi, HP 175$\,$005, India}%

Email: ajay\_singh@students.iitmandi.ac.in, samar.agnihotri@gmail.com
\thanks{Parts of this work were presented at IEEE WCNC 2016 and IEEE WCNC 2017 conferences.}

}

\maketitle
\begin{abstract}
Multicasting for disseminating popular data is an interesting solution for improving the energy and spectral efficiencies of cellular networks. To improve the achievable performance of such networks, underlay device-to-device (D2D) multicast communication offers a practical solution. However, despite significant potential for providing higher throughput and lower delay, implementing underlay D2D multicast communication poses several challenges, such as mutual interference among cellular users (CUs) and D2D multicast groups (MGs), and overhead signaling to provide channel state information, that may limit potential gains. We study a scenario where multiple D2D multicast groups may share a CU's uplink channel. We formulate an optimization problem to maximize the achievable system throughput while fulfilling quality of service (QoS) requirements of every CU and D2D MGs, subject to their corresponding maximum transmit power constraints. The formulated optimization problem is an instance of mixed integer non-linear programming (MINLP) problem, which is computationally intractable, in general. Therefore, to find a feasible solution, we propose a pragmatic two-step process of channel allocation and power allocation. In the first-step, we propose a channel allocation algorithm, which determines the subset of MGs that may share a channel subject to criteria based on two different parameters: interference and outage probabilities. Then, we propose an algorithm to allocate power to these MG subsets that maximizes the system throughput, while satisfying transmit power constraint. Numerical results show the efficacy of proposed approach in terms of higher achievable sum throughput and better spectrum efficiency with respect to various existing schemes.
\end{abstract}

\section{Introduction}
\label{Sec:1}
The unprecedented increase in the number of mobile users and a concomitant increase in their data hungry applications have created a spectrum crunch for network operators \cite{index2019trends}. However, spectrum usage measurements conducted by FCC \cite{spectrum_report} reveal that at any given time and location,  much of available spectrum remains idle. This under-utilization of spectrum has initiated numerous research efforts to propose effective spectrum management polices and techniques. In fact, 3GPP started new  standardization process for sharing spectrum in Long Term Evolution - Advanced (LTE-A) in the upcoming 5G standard \cite{kuo2015new}.

Applications such as weather forecasting, live streaming, or file distribution, which require the same chunk of data distributed to a particular group of users, naturally lend themselves to multicasting. 
Network operators may support such applications using two approaches: cellular multicast \cite{afolabi2013multicast} and mobile data offloading - shifting of local traffic to other networks \cite{andreev2015analyzing}. By exploiting the broadcast nature of 
wireless communication, cellular multicast benefits from a single unidirectional link shared among several users within the same cell. Mobile data offloading allows shifting of cellular traffic to other networks, thus offering a promising solution for offloading the core network \cite{andreev2015analyzing}. Both these approaches lead to reduced spectrum usage compared to when there is a dedicated channel for every user.

According to 3GPP \cite{chatzopoulos2019d2d}, Device-to-Device (D2D) communication is another promising technology in LTE-A to offload the burden of the centralized controller by enabling the proximate mobile users to communicate directly without going through the evolved Node Base station (eNB). In addition, it is expected to result in 
improving the energy efficiency, spectral efficiency, user experience and cell coverage \cite{araniti2017device}. Given their promise, a combination of multicasting and D2D communication appears to be an appealing solution when the same data is requested by multiple users 
in limited geographical area. Moreover, by observing the users' interest in the shared web content \cite{tatar2014survey}, such scenarios of multicasting are desirable. However, D2D multicast in cellular networks poses several intrinsic challenges, such as the data rate that can be supported is possibly decided by the user having the worst channel. Further, it is the mobile device not the Base Station (BS)\footnote{In this manuscript, the terms BS, eNB in LTE-A, and next generation NodeB (gNB) in 5G, are used interchangeably.} that transmits data, thus imposing more challenges due to limited computation and communication capabilities of mobile devices and associated reliability issues.

D2D multicast in the cellular networks can share the spectrum with cellular users in two ways: overlay and underlay communications. In overlay communication, a dedicated part of the cellular radio resources is allocated for D2D multicast. However, to support overlay communication \cite{lin2014modeling}, the channel allocation should be done judiciously to avoid under-utilization of the radio resources. In underlay communication \cite{peng2013resource}, D2D users share the cellular users' radio resources to communicate under a strict constraint to keep their interference to cellular communication below some a certain threshold. However, the mutual interference among cellular users (CUs) and D2D users may decrease or even outweigh the benefits of D2D communication. Therefore, for underlay D2D multicast optimal resource allocation schemes in presence of interference are required.

This work addresses the problem of sum throughput maximization in D2D multicast enabled underlay cellular networks. 

\subsection{Related work}
Surveys in \cite{liu2014device,mach2015band} discuss several opportunities for synergy among CUs and D2D users, where the authors evaluate the relevant state-of-art technologies, present open research challenges for enabling D2D communication, such as interference management; and novel network architectures. 

The work in  \cite{zhao2015resource} provides a tractable optimization model for maximizing the sum-rate of CUs in single cell networks. The problem of assigning communication channels in underlay D2D networks using various game-theoretic approaches is addressed in \cite{song2014game}. Furthermore, it proposes techniques for various problem formulations and solutions for different scenarios in which D2D users may compete or cooperate for resource allocation. 
Authors in \cite{lee2019performance} provide a power allocation scheme to D2D transmitters with an objective of maximizing average achievable rate, while the work in \cite{feng2018resource} provides an efficient data distribution scheme in a D2D multicast enabled cellular network by exploiting both physical and social domain factors.
In \cite{ningombam2019interference}, authors formulate the sum throughput maximization in D2D enabled cellular networks, and utilize fractional frequency to enhance the SINR of cell edge users. The work in \cite{ma2019downlink} again formulates the sum throughput maximization problem in D2D enabled cellular network, however D2D users share the downlink channels. 

A recent survey \cite{afolabi2013multicast} on multicasting in OFDMA systems provides the challenges that need to be addressed to solve twin problems of overwhelming data requirements and scarce spectrum therein. Multicast scheduling and resource allocation are inherent in two types of multicast communication: single-rate and multi-rate. In single-rate multicast communication, all the users in a particular group receive at the same rate, irrespective of their channel quality, while in multi-rate communication, the users in a group may receive at different rates as per their channel quality. The single-rate communication is further divided in three approaches: (1) predefined fixed rate - rate supported by an edge user in a cell \cite{agashe2004cdma2000}; (2) worst channel gain user rate - the group data rate is adaptively set to suit the user having the worst channel\cite{ngo2009efficient}; (3) average group throughput - data rate is based on long-term moving average throughput of the group \cite{won2009multicast}. Two approaches have been proposed to support multi-rate multicast communication: (a) information decomposition technique - splitting the multimedia data into multiple substreams and assigning a sub-carrier to each substream \cite{shao2011layered}, (b) multicast subgroup formation - splitting a group into smaller subgroups of users according to their respective channel quality \cite{zhou2014two}. Though multi-rate communication schemes are more efficient, their implementation and analysis is much more complex than for single-rate communication schemes. 

Standardization bodies such as 3GPP have also initiated efforts for enabling multicasting in 4G/5G networks \cite{3GPP_multicast}. The 3GPP has incorporated a new standard for multicast in LTE, called evolved Multimedia Broadcast Multicast Services (eMBMS) \cite{lecompte2012evolved}. eMBMS distributes content to multiple users simultaneously by using the single frequency network.   Such approaches are more effective when there is substantial temporal and spatial correlation in the requested content. In \cite{mladenov2011efficient}, the authors propose integration of application layer approaches such as forward error correction for improving the reliability when there is no-feedback from mobile users. A detailed study on dependence between data rate and eMBMS parameters is provided in \cite{monserrat2012joint}.

The popularity of video content on mobile devices has fueled research efforts to support video dissemination and to design resource allocation schemes for D2D-multicast in overlay \cite{lin2014modeling} and underlay \cite{peng2013resource,gong2015particle,meshgi2017optimal} cellular networks. In \cite{lin2014modeling}, authors provide the design, implementation, and optimization of overlay D2D-multicast, and compute various parameters, such as coverage probability of all D2D receivers, the optimal number of retransmissions for successfully delivering data packets. The work in \cite{peng2013resource} formulates a sum throughput optimization problem for underlay D2D multicast and proposes to remove those groups whose throughput is below some threshold. Authors in \cite{gong2015particle} utilize particle-swarm optimization technique for allocating the transmitted power to multicast groups (MGs), while \cite{meshgi2017optimal} considers static topology with fixed number of receivers. Recently, stochastic geometry based approaches have received considerable attention for modeling D2D communications because of their capability to accurately capture the randomness in the network geometry and provide analytical tractability to achieve precise performance evaluation \cite{wu2019network,vu2019full,bhardwaj2019d2d}. In \cite{wu2019network}, the authors leverage the stochastic geometry to model the users as Poisson Point Process, and formulate a sum throughput maximization problem in D2D integrated cellular networks. The work in \cite{vu2019full} derives closed-form expressions for coverage probability of both D2D and cellular users, and based on the derived expression, an analytical expression for D2D link sum rate is also provided. In \cite{bhardwaj2019d2d}, the authors propose a pragmatic interference management technique by considering exclusion zones around cellular users, and show that sum throughput can be further increased. However, such approaches only provide probabilistic performance guarantees for given spatial distribution. 

Most of the aforementioned works dealing with resource allocation in D2D networks consider either D2D pairs (\cite{zhao2015resource,song2014game,xu2013efficiency,wang2013energy} )
or multiple MGs with fixed number of receivers \cite{meshgi2017optimal}. To the best of our knowledge, this work for the first time considers the general problem of optimal resource allocation for multiple D2D multicasts in underlay cellular networks.

\subsection{Contributions}

This work addresses the problem of radio resource allocation in underlay communication scenarios where any number of D2D MGs may operate in a network, any MG may have any number of receivers, and any number of MGs may share the channel with a CU. Such scenarios lead to mutual interference among D2D MGs along with interference between CUs and D2D MGs sharing the channel. In order to ensure a certain level of quality-of-service to CUs and D2D multicast users, pre-defined thresholds on received signal-to-interference-plus-noise-ratio (SINR) are considered for both CUs and D2D multicast group receivers (MGRX).

The main contributions of this work are as follows:
\begin{itemize}
\item We formulate the system throughput maximization as a
Mixed Integer Non-linear Programming (MINLP) problem, subject to maximum power constraints of the CUs and D2D multicast group transmitters (MGTXs), and quality of service(QoS) constraints of both, the D2D MGRXs and CUs.
\item As MINLP problems are computationally-hard, in
general, a polynomial-time achievability scheme is proposed. This scheme operates in two stages. In the first stage, channel allocation algorithms based on two different criteria for grouping MGs are proposed for determining the set of D2D MGs which share a CU channel, and in the second stage a power allocation algorithm is proposed to maximize the sum throughput for the groups of MGs allocated to each cellular channel.
\item For two specific instances of this problem, where (1) only a single MG shares the channel with a CU, and (2) only two MGs share the channel with a CU, we provide specific schemes which are more efficient than the scheme proposed for the general case. 
\item Thorough numerical simulations of the proposed scheme are performed. Specifically, the impact of number of MGs, maximum available transmission power at CUs, QoS requirement of CUs and geographical spread of MGs on the achievable sum throughput is investigated and superiority of the proposed scheme with respect to various existing schemes is established. 
\end{itemize}

\subsection{Organization and notations} 
\noindent\textit{Organization:} Section \ref{Sec:3} introduces the system model. The problem formulation is introduced in Section \ref{Sec:4}. Section \ref{Sec:5} introduces algorithms for grouping D2D MGs sharing a particular channel based on two different criteria. Section \ref{Sec:6} provides algorithms to allocate the optimal power to MGs. It also provides power allocation schemes for two special cases where only one or two MGs share a channel. Performance evaluation of the proposed schemes is carried out in Section \ref{Sec:7}. Finally, Section \ref{Sec:8} concludes the paper.
 
\noindent\textit{Notations:} The major notation and abbreviations used in this paper are summarized in Table~\ref{Table_1_wcnc_17}. 

\renewcommand{\arraystretch}{1}
\begin{table}[htb]
\tiny
\caption{Major notation and abbreviations} 
\label{Table_1_wcnc_17}
\resizebox{\textwidth}{!}{%

\begin{tabular}[width=\linewidth]{p{0.5in} p{3.25in}}
\hline
$\mathcal{C}$ & Set of cellular users, $C = K = \vert \mathcal{C}\vert $\\
$p_{c,k}$  & The transmission power of the $k^\textrm{th}$ CU\\
$P_c^{\max}$ & The maximum power that can be transmitted by any CU\\
$P_g^{\max}$ & The maximum power that can be transmitted by any D2D MGTX\\
$\Gamma_k^c$ & The SINR received by the eNB on the $k^{\textrm{th}}$ channel\\
$h_{i,j}^k$ & The channel gain between the nodes $i$ and $j$ over the  $k$ channel\\
$R_{k,c}^{\min}$ & The minimum data rate required by the $k^{\textrm{th}}$ CU\\  
$\mathcal{G}$ & Set of multicast transmitters, G = $\vert \mathcal{G}\vert $ \\ 
$\mathcal{G}_k$ & Set of MGs communicating on channel k\\
$a_{g,k}$ & Binary variable denotes sharing of a channel\\
BS/eNB  &Base station/evolved Node Base station \\	
CU  	&Cellular user	\\
eMBMS	&evolved Multimedia Broadcast Multicast Services \\	
GDCPC	&Generalized Distributed Constrained Power Control algorithm	\\
IA-STIM	&Interference-Aware STIM	\\
LTE-A	&Long Term Evolution-Advanced	\\
MG		&Multicast Group	\\
MGRX	&Multicast Group Receiver	\\
MGTX 	&Multicast Group Transmitter	\\
MINLP	&Mixed Integer Non-Linear Programming	\\
OA-STIM	&Outage-Aware STIM	\\
QoS		&Quality of Service	\\
SINR	&Signal-to-Interference-plus-Noise Ratio \\	
SIR		&Signal-to-Interference Ratio\\ 	
STIM	&Simultaneous Transmission Interference Management 	\\

\hline 
\end{tabular}
}
\end{table}

\section{System Model}
\label{Sec:3}
This section provides the details of the network and channel models, and assumptions we adopt to design our proposed schemes to solve the joint power and channel allocation  problem for maximizing the average sum throughput.
\subsection{Network model}
We consider a D2D enabled single cell network, where CUs and D2D MGs are assumed to be uniformly distributed inside a two-dimensional circular cell of radius, $R$. Let $\mathcal{C}= \{1,2,...,C\}$ and $\mathcal{G}=\{1,2,...,G\}$ denote the set of CUs and MGs, respectively. A mobile user in the $g^\textrm{th}$ MG is denoted by $u_g$ and the set of such users is denoted by $\mathcal{U}_g$. The cardinality of $\mathcal{U}_g$ defines the number of receivers in that group: if it is one, then it models unicast communication. We assume that D2D links share the uplink channels of CUs. This assumption is justified for the following reasons: i) there is high traffic load on downlink channels \cite{mach2015band}, ii) for the uplink channels it is the base station that faces interference, however as it is 
more resourceful, it can handle it more effectively. In LTE-A systems, the smallest resource unit that can be assigned to users is a subchannel - a combination of 12 sub-carriers for one time-slot \cite{militano2015single}.
We assume that a sub-channel is used by exactly one CU, and  the cell is fully loaded. This allows us to index both CUs and orthogonal channels as $1,2,\ldots, C$. This assumption protects the CUs from strong co-channel interference from other CUs within a cell. Therefore, an uplink frequency channel $k \in \mathcal{C}$ is used by i) the $k^\textrm{th}$ CU, and ii) a set $\mathcal{G}_k\subseteq \mathcal{G}$ of $\vert\mathcal{G}_k\vert$ D2D MGs.    
  
\begin{figure}
\centering
\includegraphics[width=3.5in]{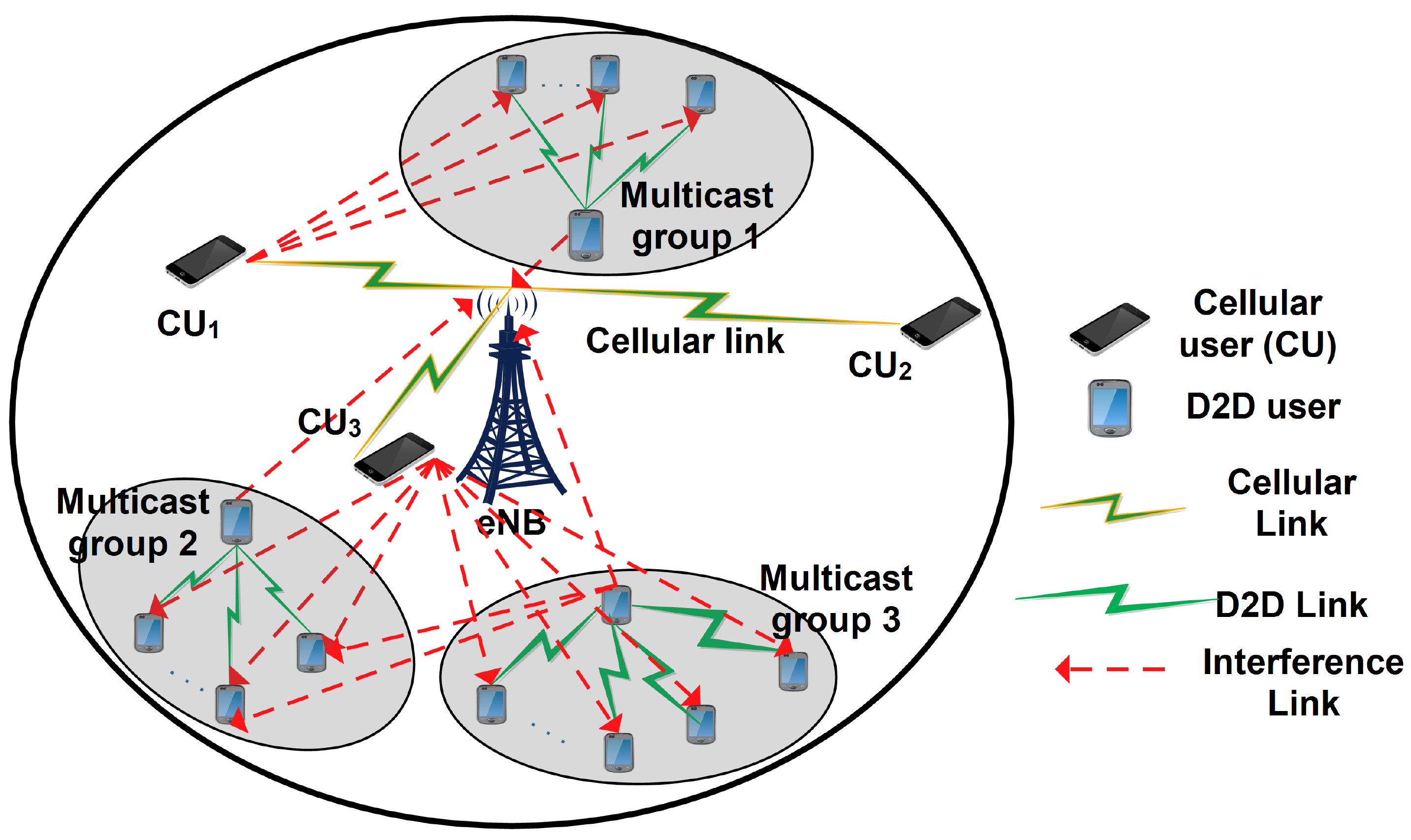}
\vspace{-0.2in}
\caption{Illustration of the D2D integrated LTE-A network. Cellular eNB, CUs and D2D users are uniformly distributed in network, the Multicast group 2 and 3 share uplink channel of the $\textrm{CU}_3$}
\label{fig:1}
\vspace{-0.2in}
\end{figure}

The work in \cite{huang2015resource} asserts that proper frequency and power allocation can mitigate the inter-cell interference. Therefore, in this work, we focus on mitigating the intra-cell interference, which arises among CUs and D2D MGs because of the channel sharing.

\subsection{Channel Model}
\label{sub_sec3:1}
The radio propagation channel model includes pathloss, shadowing, and fading, to address practical communication scenarios. The channel gain between the nodes $i$ and $j$ over the sub-carrier $s$ is given by, as in \cite{ghazzai2014optimized}: 
\begin{equation*}
h_{i,j}^s= \left(- \kappa  -  \alpha \textrm{log}_{10} d_{i,j}\right) - \xi_{i,j}^s + 10 \textrm{log}_{10} F_{i,j}^s, 
\end{equation*}
where the first factor represents the propagation loss: $d_{i,j}$ denotes the distance between the nodes $i$ and $j$, $\kappa$ is the pathloss constant, and $\alpha$ is the path loss exponent whose value lies between 2-6; the second factor  $\xi_{i,j}^s$ represent the log-normal shadowing with zero mean and variance $\sigma_\xi^2$; and the last factor $F_{i,j}^s$ accounts for Rayleigh fading (typically considered with a Rayleigh parameter $\delta$ such that $E\left[\delta^2\right] = 1$). A block fading model is considered, where the fading remains constant over the sub-carriers of a resource block.
 
\subsection{Achievable Throughput}
We assume that to maximize the spectral efficiency a frequency channel is shared by multiple D2D MGs and a CU. During an uplink time slot, the multicast group transmitters (MGTXs) may interfere with the co-channel CU signal, at the eNB receiver. Similarly, the transmitter of CU and other co-channel D2D MGs may interfere at the receivers of a D2D MG. Specifically, in a particular uplink time slot, let $s_k$ and $s_g$ denote the unit-variance signal transmitted by the $k^\textrm{th} \left(k \in \mathcal{C}\right)$ CU and the $g^\textrm{th}\left( g \in \mathcal{G} \right)$ MGTX, respectively. Let $\mathcal{G}_k$ denote the set of D2D MGs sharing the channel with the $k^{\textrm{th}}$ CU. Therefore, signal received at the eNB and the $r^{\textrm{th}} \left(r \in \mathcal{U}_g\right) $ MGRX on the $k^{\textrm{th}}$ channel can be written, respectively, as
\begin{align*}
y_{c}^k&= h_{c,b}^k \sqrt{p_{c,k}} s_k + \sum_{g=1}^{\vert \mathcal{G}_k \vert} \sqrt{p_{g,k}} h_{g,b}^{k} s_g + \mathcal{Z}_c\\
y_{r}^k&= h_{g,r}^k \sqrt{p_{g,k}} s_g + \sum_{j=1,j\neq g}^{\vert \mathcal{G}_k \vert} \sqrt{p_{j,k}} h_{j,r}^{k,g} s_j+ \sqrt{p_{c,k}} h_{c,r}^k s_k + \mathcal{Z}_d,
\end{align*}
where $p_{c,k}$ and $p_{g,k}$ denote power transmitted by the $k^\textrm{th}$ CU and the $g^\textrm{th}$ MGTX, respectively; $h_{c,b}^k$, $h_{g,b}^k$, $h_{g,r}^k$ and $h_{c,r}^k$ denote the channel gains between CU - eNB, MGTX - eNB, MGTX - MGRX and CU - MGRX, all communicating on the $k^\textrm{th}$ channel, respectively. Variables $\mathcal{Z}_c$ and $\mathcal{Z}_d$ denote the additive white Gaussian noise with zero mean and variance $N_0$. 
The achievable SINR and data rate corresponding to the CU that communicates on the channel $k$ with the transmit power $p_{c,k}$ are given, respectively, by
\begin{align}
&\Gamma_c^{k}\left(k,p_{c,k}\right) = \left( \dfrac{p_{c,k} h_{c,b}^k}{\sum_{g=1}^{\vert \mathcal{G}_k\vert} p_{g,k} h_{g,b}^k  + N_0}  \right) \nonumber\\
&R^c_{k}\left(k,p_{c,k}\right)=  B_k \log_2 \left( 1+ \Gamma_c^{k} \right) \label{eq:4a},
\end{align}
where $B_k$ denotes the bandwidth of the $k^{\textrm{th}}$ channel. Therefore, the achievable rate of all the CUs can be expressed as follows: 
\begin{equation}
\label{ch_2_eq:cu_opt}
R_{c} = \sum_{k=1}^{\vert \mathcal{C} \vert}  R_{k}^c 
\end{equation}  

As per the system model, the $g^{\textrm{th}}$ MG shares resources with the $k^{\textrm{th}}$ CU and with other co-channel MGs. Therefore, the CU and such MGs create interference at the receivers of the $g^{\textrm{th}}$ MG. Thus, SINR at the $r_g^\textrm{th}$ receiver (the $r^\textrm{th}$ receiver belonging to the $g^\textrm{th}$ MG) can be expressed as follow:
\begin{equation*}
\label{eq:5}
\Gamma_{r,g}^{k} = \dfrac{p_{g,k} h_{g,r}^{k}}{\sum_{j=1,j\neq g}^{\vert \mathcal{G}_k \vert} p_{j,k} h_{j,r}^{k,g} + p_{c,k} h_{c,r}^k + N_0},
\end{equation*}
where $h_{g,r}^{k}$ denotes the channel gain from the $g^{\textrm{th}}$ MGTX to its $r^{\textrm{th}}$ 
receiver on the $k^\textrm{th}$ channel, $h_{c,r}^k$ is interference 
channel gain from the $k^{\textrm{th}}$ CU to the $r^{\textrm{th}}$ 
receiver, $h_{j,r}^{k,g}$ denotes the interference gain from the $j^{\textrm{th}}$ MGTX  to the $g^\textrm{th}$ MG's $r^{\textrm{th}}$ D2D receiver.
The maximum achievable rate in a MG is determined by the SINR of the worst user. Therefore, the achievable SINR and data rate corresponding to the $g^{\textrm{th}}$ MG that communicates on the $k^{\textrm{th}}$ channel are given, respectively, by
\begin{align}
\Gamma_{g}^k &= \min_{r\in \mathcal{U}_g}{\Gamma_{r,g}^{k}} \nonumber\\
R^{D}_{k,g} &= B_k \log_{2}\left(1+ \Gamma_{g}^k\right),\label{eq:7_ch1}
\end{align}
The achievable data rate for the $k^\textrm{th}$ CU without channel sharing can be determined by 
\begin{equation}
\label{eq:add_1}
\widehat{R^c_k} = B_k \log_{2} \left( 1 + \dfrac{p_{c,k} h_{c,b}^k}{N_0}\right)
\end{equation} 
From \eqref{eq:4a} and \eqref{eq:add_1}, it can be observed that there is a reduction in the achievable data rate of the $k^{\textrm{th}}$ CU, and the reason for this is the interference caused by the co-channel MGs. Thus, the data rate reduction for any CU can be determined as 
\begin{equation*}
\label{eq:add_2}
\Delta R_{k}^c = \widehat{R^c_{k}} - R^c_{k}
\end{equation*}
Therefore, the achievable throughput gain when D2D MGs in $\mathcal{G}_k$ share resources with the $k^\textrm{th}$ CU can be expressed as
\begin{equation}
\label{eq:add_3}
\Delta R_{k,g}= \sum_{g \in \mathcal{G}_{k}}R^D_{k,g} + R^c_{k} - \widehat{R^c_{k}}  = \sum_{g \in \mathcal{G}_{k}}R^D_{k,g} - \Delta R^c_{k},
\end{equation} 

\section{Problem Formulation}
\label{Sec:4}
In this section, we formulate the sum throughput maximization problem in multiple D2D multicasts enabled underlay cellular networks. We first introduce the objective function, along with constraints, then we discuss the MINLP model for the sum throughput maximization problem in a hybrid CU and D2D multicast wireless networks. 

The objective of the resource allocation problem in a hybrid network (i.e CU and D2D multicast network) is to allocate the channel and power to the transmitters of the MGs that maximize the aggregate sum throughput. Let binary variable $a_{g,k}, g \in \mathcal{G}, k \in \mathcal{C}$ denote the temporal assignment of the $k^\textrm{th}$ CU channel to the $g^\textrm{th}$ D2D MG. Specifically, $a_{g,k} = 1$ indicates that the $g^\textrm{th}$ MG shares the channel with the $k^\textrm{th}$ CU, and $a_{g,k} = 0$ indicates otherwise.

The achievable rate of all the MGs can be expressed as follows: 
\begin{equation}
\label{ch_2_eq:9}
R_{g} = \sum_{k=1}^{\vert \mathcal{C} \vert} \sum_{g=1}^{\vert \mathcal{G} \vert} a_{g,k}  R_{k,g}^D 
\end{equation}  

To ensure a minimum QoS level to CUs, we assume, as in \cite{hasan2014resource}, that there is a threshold on minimum acceptable data rate, $R_{c,k}^{\min}$ for CUs. In other words, there is a limit for maximum interference that can be accepted over each resource block (RB) for each CU, that is

\begin{equation}
\label{thres_Ith}
R_{k,g}^c = B_k \log_2 \left( 1+ \dfrac{p_{c,k} h_{c,b}^k}{N_0 +   \sum_{g=1}^{\vert \mathcal{G}\vert} a_{g,k} p_{g,k} h_{g,b}^k}\right) \geq R_{c,k}^{\min}
\end{equation}
By rearranging, the minimum power threshold on each CU can be defined as
\begin{align}
P_c^{\min} = \frac{(2^{R_{c,k}^{\min} /B_k} - 1) (N_0 +  \sum_{g=1}^{\vert \mathcal{G}\vert} a_{g,k} p_{g,k} h_{g,b}^k)}{h_{c,b}^k} \label{eq:P_C_min}
\end{align}

The maximum power that can be assigned to any user is also limited, that is $P_c^{\max}$ for a CU and $P_g^{\max}$ for each MGTX. Since the co-channel MGTXs may decrease the CU's SINR, to ensure the QoS, each CU should have a minimum power constraint $P_c^{\min}$, whose value can be computed from \eqref{eq:P_C_min}. By utilizing expressions \eqref{ch_2_eq:cu_opt} and \eqref{ch_2_eq:9}, the overall system throughput maximization problem can be stated formally as
\begin{align}
\mathbf{P1:}~~~ &\underset{a_{g,k}, p_{g,k},p_{c,k}} {\max~}
\left( R_c + R_g \right) \nonumber\\
\text{s.t.} \quad & \text{C}_1:P_c^{\min} \leq p_{c,k} \leq P_c^{\max}, \forall c \in \mathcal{C} \nonumber\\
& \text{C}_2: 0 \leq p_{g,k} \leq P_{g}^{\max}, \forall k \in \mathcal{K}  \nonumber\\ 
& \text{C}_3: \Gamma_{g}^k \geq \gamma_{r}^{\textrm{th}}, \Gamma_c^{k} \geq \gamma_{k}^{\textrm{th}} \nonumber\\
&\text{C}_4: \sum_{g\in \mathcal{G}}a_{g,k} = G_k, \forall g \in \mathcal{G} \nonumber\\
& \text{C}_5: a_{g,k} \in \{0,1\}, g \in \mathcal{G}, k \in \mathcal{K}, \label{main_problem}
\end{align}
where the constraints $\text{C}_1$ and $\text{C}_2$ limit the maximum power allocated to D2D MGTXs and CUs, respectively. Constraint $\text{C}_3$ guarantees minimum achievable rate for every CU and every receiver of every MG. The constraint $\text{C}_4$ limits the number of MGs per channel to $G_k$, while constraint $\text{C}_5$ indicates whether the $g^{\textrm{th}}$ MG shares the resources with the $k^\textrm{th}$ CU or not.
The throughput expressions of the CU and D2D MGs (equations  \eqref{ch_2_eq:cu_opt} and \eqref{ch_2_eq:9}) show that allocation of different channels and powers to different users has different contributions to the throughput gain. Therefore, to maximize the system throughput we need to determine the optimal set of CUs and D2D MGs that share a channel and optimally allocate powers to them.

The formulation in Problem P1 is an instance of MINLP problem, since some decision variable are integers (more specifically, $a_{g,k}$), and other variable are continuous ($p_{c,k},p_{g,k}$). Such problem are known to be  NP-hard, in general \cite{kannan1978computational}. Even if we relax the integer constraint, $a_{g,k}$, it is still difficult to obtain the globally optimal solution, as the relaxed form is non-concave in both $p_{c,k}$ and $p_{g,k}$. Therefore, to solve the resource allocation problem efficiently, we propose a pragmatic approach that divides the problem in two subproblems: 1) construct the subsets of the multicast groups which share the channel with a CU (the channel allocation problem), 2) allocate powers to the CUs and the transmitters of co-channel MG subsets thus constructed to maximize the overall throughput (the power allocation problem.) 

The following proposition provides a lower bound to the optimal solution of Problem P1. 

\begin{pavikp}
\label{proposition:1}
For any value of $p_{g,k}, p_{c,k}$ and channel gains, we have 
\begin{align*}
\label{eq:11_wcnc_17}
\sum_{k=1}^{\vert \mathcal{K} \vert} R^c_{k,g}{\left(k,p_{c,k}\right)} > 
\sum_{k=1}^{\vert \mathcal{K} \vert}  \log_2\left(p_{c,k} h_{c,b}^k\right) - \sum_{k=1}^{\vert \mathcal{K} \vert} \sum_{g=1}^{\vert \mathcal{G}_k\vert }  \log_2(P_g^{\max}h_{g,b}^k) 
\end{align*}
\label{Prop_1}
\end{pavikp}

\begin{IEEEproof}
Please refer to Appendix ~\ref{appendix_proposition}
\end{IEEEproof}

The above mentioned proposition shows a lower bound on the maximum throughput is obtained when all D2D MGTXs transmit at the maximum power $(P_{g}^{\max})$. Thus, for all realizations of channel gains, the accuracy of the bounds depends only upon the maximum value of transmitter power of MGTXs. 

\section{Channel Allocation}
\label{Sec:5}
To solve the channel allocation problem, we provide two approaches: 1)  interference-aware channel allocation, and 2) outage-aware channel allocation. 

The interference-aware channel allocation scheme insures that the MGs which have strong mutual interference must not share a channel, otherwise their contribution to the sum throughput may be very low. Furthermore, as a practical matter, different mobile users depending upon their running applications may have different QoS requirements. Therefore, we also propose a channel allocation scheme which allocates the channel based on the specific QoS requirement of D2D users. 
Specifically, the QoS problem is tackled by proposing three channel allocation schemes based on the outage probabilities of D2D MGs. The motivation behind doing so is that, in channel sharing schemes CU faces interference from MGTXs, and as long as its target rate (depending on its QoS) is achieved, it does not care if the channel is shared or not. The same holds true for each MGRX which faces interference from the CU and other MGTXs sharing the channel with it.
\subsection{Interference-aware Channel Allocation}
\label{Sec:5_wcnc_17}
In this subsection, we address the problem of determining the set of D2D MGs which may share the CU channel based on a criterion to minimize mutual interference among the MGs sharing a channel. 

The MGs that share channel with the $i^\textrm{th}$ CU can be deemed as a single group, and denoted as  $\mathcal{G}_i, i = 1,
\cdots, C$. We present an algorithm for grouping of D2D MGs that share the channel with a CU. The D2D MGs which contribute positively to the system throughput while sharing resources with a CU, are grouped together. However, we eliminate those groups which cause severe mutual interference. The reason for this is that even if two groups individually contribute positively to the system throughput, their mutual interference may be so strong that these groups do not find a suitable CU to share a channel with. Therefore, by putting a threshold ($\gamma_{th}$) on the ratio of their respective channel gains, we can prevent such MGs from sharing a channel. The throughput gained obtained when multiple D2D MGs share a channel with a CU can be obtained by Equation \eqref{eq:add_3}. For the channel allocation part, we assume that all MGTXs transmit at the maximum transmission power. However, after channel allocation, in power allocation step, transmit powers for co-channel MGs are adjusted such that the sum throughput is maximized. Let $h_{j,j'}^{k}$ denote the channel gain from the $j^{\textrm{th}}$ MG to the receivers of co-channel $j^{'\textrm{th}}$ MG, when both transmit on the $k^\textrm{th}$ channel. On every channel, we find the throughput gain achieved by every MG using \eqref{eq:add_3}. With these insights, we design an algorithm for the channel allocation. A formal description of the interference-aware channel allocation algorithm is provided in Algorithm~ \ref{algo_wcnc_17}. Its major steps are as follows: 
\begin{enumerate}
\item First (line 4), for each channel, sort the MGs in decreasing order of their contribution to the sum throughput.
\item In Step 2, (lines 5-7), only those MGs are allowed to share the channel which provide positive throughput gain.
\item Next, (lines 8-13), the ratio between the channel gains of the MGs which contributed positively to the sum throughput is calculated. 
The MGs which are in close proximity of each other have lower channel ratio values, and may not be available to share the same channel because of high mutual interference. Therefore, those can be prevented by putting a threshold $\gamma_{th}$ on the channel gain ratio. 
\end{enumerate}

\begin{algorithm}[t]
 \caption{Interference-aware channel allocation to CU and D2D MGs}
 \DontPrintSemicolon
\Begin($ $)
{
\label{algo_wcnc_17}
$\textrm{Initialize array} ~ \mathcal{G}^{'}_k = \phi, \mathcal{G}_k = \phi,  \forall k= 1,\ldots, C$ and $G_j \left(j=1, \ldots, G\right)$ \;
Evaluate $\Delta R_{g,k}$, $\forall k \in \mathcal{C}$ and 
$g \in \mathcal{G}$;
\For{k = 1 to $C$}{
sort the $\Delta R_{g,k}$ in descending order\\
\For{\textrm{g = 1 to G}}{
\If{$\Delta R_{g,k} > 0$}
{
 update set $\mathcal{G}^{'}_k = \mathcal{G}^{'}_k \cup g$  
}
}
}
\For{$\forall k \in \mathcal{C}$}
{
j = $\mathop{\text{argmax}}_{j \in \mathcal{G}^{'}_k} \Delta R_{g,k}$\;
$\mathcal{G}^{'}_k =  \mathcal{G}^{'}_k \setminus j$\;
$\mathcal{V}_1 = \dfrac{h_{j,r}^{k,g}}{h_{j',j}^{k}}$, $\mathcal{V}_2 = \dfrac{h_{j',r'}^{k}}{h_{j,j'}^{k}}$ $\forall j' \in  \mathcal{G}^{'}_i $;

\If{$\mathcal{V}_1 > \gamma_{th}$ and $\mathcal{V}_2 > \gamma_{th}$}{
$\mathcal{G}_i = \mathcal{G}_i \cup j$, and $\mathcal{G}_j = \mathcal{G}_j \cup i$
}
}
}
\end{algorithm}

\subsection{Outage-aware Channel Allocation}
\label{Sec:3_5}
In this subsection, we provide another approach for channel allocation by minimizing outage probabilities of D2D MGs. Similar to interference-aware channel allocation, let the MGs that share channel with the $i^\textrm{th}$ CU are deemed as a single group, and denoted as  $\mathcal{G}_i, i = 1,\cdots, C$. As per the system model any number of D2D MGs may share the channel with a CU, while maintaining QoS requirement of the CU. Let $G_k$ be the number of MGs that share the channel with the $k^\textrm{th}$ CU. 

Before providing a solution to the outage-aware channel allocation problem, we provide a general form of outage probability of a D2D receiver.  
An outage event for MG `g' occurs if the SIR ($\Gamma_{r,g}^{k}$) of the $r_g$ receiver falls below its target SIR $\gamma_\textrm{th}^{o}$. Assuming that MGTX transmits at maximum power, that is, $p_{g,k}$, the outage probability of the $g^\textrm{th}$ MG is given by the following lemma.
\begin{pavikl}
\label{lemma_op}
The outage probability of a D2D receiver is 
\begin{equation}
\label{eq_prob}
P_{out}^{G_{k,i}} = \text{Pr} \left(\Gamma_{r,g}^{k} \leq \gamma_{\textrm{th}}^{o} \right) = 1 - \exp{\left(- \chi_{g,k} {\gamma_\textrm{th}^{o}}^{\frac{2}{\alpha}} d_{g,r}^2 \left[ \lambda_c \left( \frac{p_{c,k}}{p_{g,k}} \right)^{\frac{2}{\alpha}} + \lambda_g \right]\right)}
\end{equation}
where $ \chi_{g,k} = \pi \Gamma \left(1 + \frac{2}{\alpha} \right) \Gamma \left(1 - \frac{2}{\alpha} \right)$,  $d_{g,r}$ is radius of D2D MG, and $\Gamma\left(x\right)$ denotes Gamma function.  
\end{pavikl}
\begin{IEEEproof}
Please refer to Appendix~\ref{proof_OP}
\end{IEEEproof}

By utilizing this outage probability, we present three objectives for selecting a CU. Initially, it is assumed that MGTXs transmit at maximum power, $P_{g}^{\max}$.

Objective 1: Choose the CU that minimizes the outage probability  of any particular group, such as for the $j^\textrm{th}$ MG.
\begin{equation}
\label{eq:28}
i^*= \argmin_{i \in \mathcal{C}} P_{out}^{G_{j,i}} \nonumber
\end{equation}

Objective 2: Choose the CU that minimizes the maximum of outage probabilities of $C$ D2D MGs as:
\begin{equation}
\label{eq:29}
i^*= \argmin_{i \in \mathcal{C}} \max_{\substack{G_1,\ldots,G_k \in {\mathcal{G}}}} 
\left( P_{out}^{G_{1,i}},...,P_{out}^{G_{K,i}}\right) \nonumber
\end{equation}

Objective 3: Choose the CU that minimizes the sum of outage probabilities of $\vert \mathcal{G}_k \vert$ D2D MGs.
\begin{equation}
\label{eq:30}
i^*= \argmin_{i \in \mathcal{C}}  \sum_{j=1}^{\vert \mathcal{G}_k \vert} P_{out}^{G_{j}}\nonumber
\end{equation}

Depending on the requirement of application running on each D2D MG, the BS may use the optimal objective for selecting the CU. For example, if the $j^\textrm{th}$ group has more stringent QoS requirements than other groups, the base station opts for Objective 1 to guarantee performance for the $j^\textrm{th}$ group. Objective 2 tries to minimize the maximum of outage probabilities of the $G_k$ D2D groups, thus it attempts to maintain the SIR of weakest group above some threshold value. Objective 3 tries to minimize the sum of outage probabilities of $G_k$ groups, thus it tries to maintain fairness among all 
groups. However, as we increase the number of groups sharing the channel with a CU, computing the conditional outage probability gets increasingly cumbersome. One way to reduce this computational effort is to consider the interference only from the dominating D2D MG (which causes the highest interference). A formal description of the proposed outage-aware channel allocation algorithm is provided in Algorithm~\ref{chap_3_channel_allocation}.

\begin{algorithm}[t]
\caption{Outage-aware channel allocation to CU and D2D MGs}
\label{chap_3_channel_allocation}
\DontPrintSemicolon
\KwIn{ Initialize array $\mathcal{G}^{'}_k = \phi, \mathcal{G}_k = \phi, \mathcal{G}_k = \phi,  \forall k= 1,\ldots, N$ and $G_j \left(j=1, \ldots, G \right)$ \\
 Evaluate outage probability, $P_{out}^{g,k}$, $\forall k \in \mathcal{C}$ and 
$g \in \mathcal{G}$ using \eqref{eq_prob}.
}
\Begin
{
\For {k = 1 to $C$} 
{
sort $P_{out}^{g,k}$ in ascending order. \\
\For {g = 1 to G}
{
\If{$P_{out}^{g,k} < \Theta_g^{\textrm{th}}$ }
{
update set $\mathcal{G}^{'}_k  = \mathcal{G}^{'}_k \cup g$ \\
}}}
\For {k = 1 to $C$} 
{
\For {j = 1 to  $\mathcal{G}^{'}_k$}
{ 
i = $\mathop{\text{argmin}}_{k \in \mathcal{C}}$ 
\{\textrm{Objective 1 or Objective 2 or Objective 3}\} \\
 $\mathcal{G}^{'}_k =  \mathcal{G}^{'}_k \setminus k$ \\
$\mathcal{V}_1 = \sum_{g=1}^{\mathcal{G}_k} p_{g,k} h_{g,b}^k$\\

\If {$\mathcal{V}_1 \leq I_{th}^k$}
{
$\mathcal{G}_k = \mathcal{G}_k \cup j$, and $\mathcal{G}_k = \mathcal{G}_k \cup k$}
}}
}
\end{algorithm}
\section{Power allocation}
\label{Sec:6}
After assignment of different subsets of MGs to different channels,  the integer constraint $a_{g,k}$ vanishes from \eqref{main_problem}. Now, we need to allocate the optimal transmit powers to the MGs sharing a channel to maximize the sum throughput. Before solving the general problem of power allocation, we address two special cases: a single MG shares the channel with a CU, that is $G_k = 1$; and two MGs share the channel with one CU, that is $G_k = 2$. For these cases, more efficient specific solutions can be constructed than that for the general case, as discussed below.

\subsection{Power allocation for a single MG on a channel}
This is a special case of Problem P1, when a single MG shares the channel with a CU. The main optimization Problem P1 can be divided into two subproblems - channel allocation and transmit power allocation. The channel allocation subproblem can be solved using bipartite graph based matching algorithm \cite{hopcroft1973n}, and the power allocation subproblem can be solved using Lemma \ref{ch_2_lemma_1} below.

In bipartite graph based matching, all D2D MGs and the CUs are partitioned into two groups of vertices of a graph and edge weight between two vertices such as between the $k^\textrm{th}$ CU and the $g^\textrm{th}$ D2D MG is denoted by $R_{k,g}$. To calculate $R_{k,g}$ initially all CUs and MGTX are assumed to  transmit at the maximum power. Hence, the problem of finding an optimal channel allocation for maximization of sum throughput can be written as:
\begin{subequations}
\begin{align}
\label{cu_mg}
\mathbf{P2:}~~~ &\mathop{\text{arg max} }_{k \in \mathcal{C}} \left( \sum_{g=1}^{G} R_{k,g}\right) \\
&\text{such that \ } C \geq G  \label{cu_mg_1}\\ 
&\sum_{g \in \mathcal{G}} a_{g,k} = 1, \textrm{for}~ \forall k \in \mathcal{C} \label{cu_mg_2} 
\end{align}
\end{subequations}
Constraints \eqref{cu_mg_1} - \eqref{cu_mg_2} imply that number of CUs must be greater than or equal to number of MGs for optimal matching and a channel is not to be shared by more than one MG. After calculating the edge weight matrix, the bipartite graph based optimal matching problem can be solved by well known Hungarian algorithm \cite{kuhn2005hungarian} with time complexity $O \left(G C^2\right)$. 

Next, we derive the CUs' and D2D users' optimal transmitting power. As mentioned earlier, the maximum power that can be assigned to any user is limited, that is $P^{\max}_{c}$ for CU and $P^{\max}_{g}$ for D2D MG, respectively. To find the optimal powers ($p_{c,k}^{*},  p_{g,k}^{*}$), we prove the following lemmas.

We assume that all the gains $h_{c,b}^k$ and $h_{g,r}^k \geq 0$ because if any of the communication link between a CU andthe  BS, or co-channel MGTX to its worst receiver is blocked, then the sum rate for that link is zero and does not depend upon the power allocated. Therefore, the optimal value of $\left(p_{c,k}^{*}, p_{g,k}^{*}\right)$ is $\left(0,P^{\max}_{g}\right)$ or $\left(P^{\max}_{c},0\right)$, respectively.

\begin{pavikl}
\label{ch_2_lemma_1}
The optimal value of power is $\left(P_{i}^{*},P_{j}^{*}\right)= \left( P_{i}^c= P_c^{\max}\right)\text{ or } \left( P_{j}^D= P^{\max}_{g} \right)$.
\end{pavikl}
\begin{IEEEproof}
Please refer to Appendix~\ref{appndx2:lemma_1}
\end{IEEEproof}
\begin{pavikl}
\label{ch_2_lemma_2}
The optimal transmit power allocation $\left(P_{i}^*,P_{j}^*\right)$ over a multicast group and CU only exists on the corners of feasible region.
\end{pavikl}
\begin{IEEEproof}
Please refer to Appendix~\ref{appndx2:lemma_2}
\end{IEEEproof}

The example in Appendix \ref{appndx:examples_1} demonstrates the use of Maximum Weighted Bipartite Matching for optimally solving the power allocation problem for $G_k$ =1. 

In the next subsection, we address the scenarios where the number of MGs per channel is two.
\subsection{Power allocation for two MGs on a channel}
From interference-aware and outage-aware channel allocation algorithm, we know the MGs that share the channel with a CU. When two MGs share the channel with a CU, the optimization problem is reduced to determining the optimal transmit powers that maximize the sum throughput, while satisfying the rate requirement of every CU and MG. The SINR expressions for MGs and CU can be written as
\begin{align}
\Gamma_{r,g_1}^k &= \dfrac{p_{g_1,k} h_{g_1,r}^k}{p_{g_2,k}h_{g_2,r}^{k} + p_{c,k} h_{c,r}^k + N_0} \geq \gamma_{r}^\textrm{th}\label{subsection_eq_2}\\
\Gamma_{r',g_2}^k &= \dfrac{p_{g_2,k} h_{g_2,r'}^k}{p_{g_1,k} h_{g_1,r'}^{k} + p_{c,k} h_{c,r'}^k + N_0} \geq \gamma_{r'}^{\textrm{th}} \\
\Gamma_{c}^k &= \dfrac{p_{c,k} h_{c,b}^k}{p_{g_1,k} h_{g_1,b}^{k} + p_{g_2,k} h_{g_2,b}^k + N_0} \geq \gamma_{c}^{\textrm{th}} \label{subsection_eq_2_1}
\end{align}
\begin{equation}
\left(p_{g_1,r} , p_{g_2,r'}\right) \leq P^{\max}_g \text{and } p_{c,k} \leq P^{\max}_c, \label{subsection_eq_4}
\end{equation}
where $\gamma_c^\textrm{th}, \gamma_{r}^\textrm{th}, r \in g_1$; and $\gamma_{r'}^\textrm{th}, r' \in g_2$ denote the minimum SINR threshold of CU, $MG_1$ and $MG_2$, respectively; and \eqref{subsection_eq_4} states the maximum power constraint. 
By rearranging the inequalities \eqref{subsection_eq_2}-\eqref{subsection_eq_2_1}, we have:
\begin{subequations}
\begin{align}
f_{g_1} &\triangleq  p_{g_1,k} h_{g_1,r}^k - \gamma_{r}^\textrm{th} p_{g_2,k} h_{g_2,r}^k - \gamma_{r}^\textrm{th} p_{c,k} h_{c,r}^k - \gamma_{r}^\textrm{th} N_0 \geq 0,\label{subsection_eq_5}\\
f_{g_2} &\triangleq  p_{g_2,k} h_{g_2,r'}^k - \gamma_{r'}^\textrm{th} p_{g_1,k}h_{g_1,r'} - \gamma_{r'}^\textrm{th} p_{c,k} h_{c,r'}^k - \gamma_{r'}^\textrm{th} N_0 \geq 0,\label{subsection_eq_6}\\
f_{c} &\triangleq p_{c,k} h_{c,b}^k - \gamma_{c}^\textrm{th} p_{g_1,k} h_{g_1,b}^{k} - \gamma_{c}^\textrm{th}  p_{g_2,k} h_{g_2,b}^k - \gamma_{c}^\textrm{th} N_0 \geq 0,\label{subsection_eq_7}
\end{align}
\end{subequations}

\begin{figure}[t]
\centering
\includegraphics[width=3.5in]{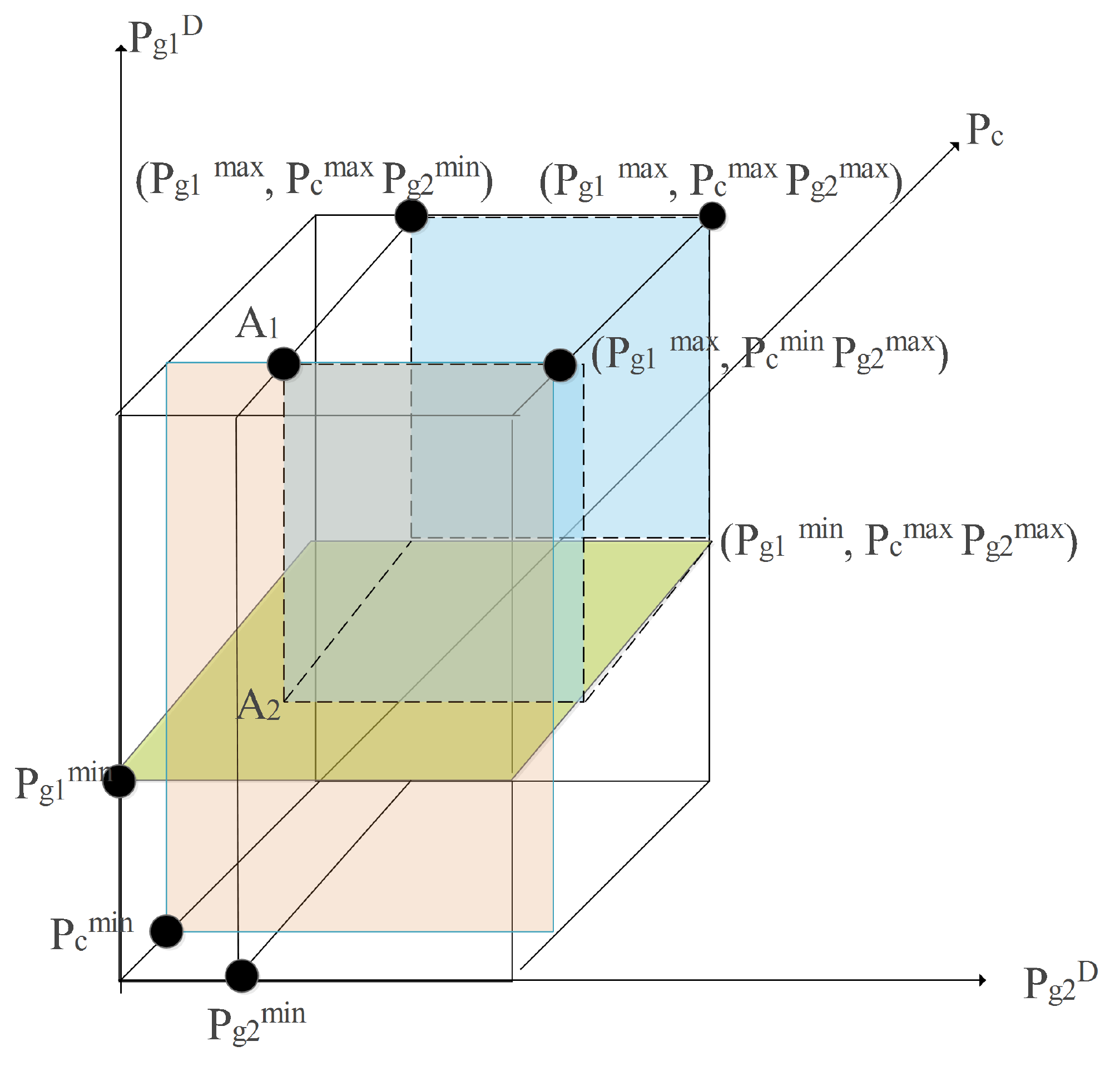}
\caption{Power region when two multicast group shares the resources with one cellular user.}
\label{fig:2_cubes}
\end{figure}

The system parameter are set in such a way that, the augmented matrix associated with \eqref{subsection_eq_2}-\eqref{subsection_eq_2_1} is full rank. This condition assures that the power planes interest at some point and a unique solution to power allocation problem can be obtained.
Equations \eqref{subsection_eq_5}-\eqref{subsection_eq_7} show that relations among the SINR thresholds and the transmit powers can be characterized as planes in a 3-dimensional space as depicted in Figure \ref{fig:2_cubes}. 

The minimum power constraint of every mobile user defines the point of intersection of the corresponding plane with a power axis. The 3D upper right corner region within a cube depicts the linear inequalities \eqref{subsection_eq_5}-\eqref{subsection_eq_7}.
The surfaces, \eqref{subsection_eq_5}-\eqref{subsection_eq_7}, of the cube depict the maximum individual power constraints. For the sake of simplicity, we omit the subscript $k$, as $k$ can be any channel. 
To define the minimum power that needs to be allocated to satisfy the SINRs thresholds of CU, $MG_1$ and $MG_2$, we assume that there is no mutual interference between any pair of D2D MGs, and between CU and D2D MGs. Therefore, we have $p_c^{\min}= \dfrac{\gamma_c^\textrm{th}N_0}{h_{c,b}}$, $ p_{g_1}^{\min} = \dfrac{\gamma_{r}^\textrm{th} N_0}{h_{g_1,r}}$ $p_{g_2}^{min} =\dfrac{\gamma_{r'}^\textrm{th} N_0}{h_{g_2,r'}}$, and the maximum power that can be allocated is 
$p_c= P_c^{\max}$, $ p_{g_1} = P_{g_1}^{\max}$, $p_{g_2} = P_{g_2}^{\max}$. 
The feasible power region is 
\begin{equation*}
\left(p_c^{\min} \leq p_c \leq P_c^{\max} \right),
\left( p_{g_1}^{\min} \leq p_{g_1} \leq p_{g_1}^{\max}\right),  
 \left(p_{g_2}^{\min} \leq p_{g_2} \leq p_{g_2}^{\max}\right), \label{power_min_th_3}
\end{equation*}
where $p_c^{\min}, p_{g_1}^{\min}, p_{g_2}^{\min}$
are the minimum powers that fulfill the individual rates. However, these powers only satisfy the constraints, and may not maximize the system throughput. The coordinates of point $A_1$ (in Figure \ref{fig:2_cubes}) are positive because it is an intersection of power planes. The feasible power region is shaped by the overlapping of three planes and by three faces of the cube. 
\begin{pavikl}
\label{ch_2_lemma1_MG_2_1}
The optimal transmit power allocation $\left(p_{c,k}^*, p_{g_1,k}^*, p_{g_2,k}^*\right)$ over a CU and MGs only exists on the surface of permissible power region.
\end{pavikl}
\begin{IEEEproof}
Please refer to Appendix ~\ref{appendices:lemma1}
\end{IEEEproof}

However, searching for the optimal solution on a surface by an exhaustive search is computationally inefficient. Therefore, to solve it efficiently, an approximate 
solution can be constructed by searching only the vertices of the permissible region, which we will prove as follows. 
The above lemma says that the optimal power lies on the surfaces of the  permissible power region. Therefore, we need to prove that objective function is quasi-convex, which ensures that global maximum lies only on the corner points. 
\subsubsection*{Power allocation based on corner search method for $G_k=2$}

We present the power allocation for the CUs and MGTXs' to maximize the sum rate when only two MGs share a channel with a CU. 
To find the optimal values of $p_{c,k}$, $p_{g_1,k}$ and $p_{g_2,k}$, an assumption is made that all the gains $h_{c,b}^k$ and $h_{g1,r}^k,h_{g2,r'}^k \geq 0$, because otherwise the optimal solution $\left(p_{c,k}^*, p_{g_1,k}^*,p_{g_2,k}^*\right)$ is $\left(0,P^{\max}_{g},0\right)$, $\left(0,0,P^{\max}_{g}\right)$ or $\left(P^{\max}_{c},0,0\right)$.
We now propose two lemmas to show that the optimal transmit power lies on only on the surface of the permissible power region, and maximum values lies only at the corner points.  

\begin{pavikl}
\label{ch_2_lemma1_MG_2_2}
The sum-rate function is quasi-convex on a boundary, ensuring that the maximum values are at the corners.
\end{pavikl}
\begin{IEEEproof}
Please refer to Appendix ~\ref{ch2_appendices_2}
\end{IEEEproof}

From the above lemma, it can be inferred that an approximate solution of power allocation problem lies at the corner points of permissible power region. It is also concluded that either the D2D MG or CU will transmit at the maximum power for maximizing the system throughput. 
Let $\lbrace p_{g_1},p_{g_2}\rbrace_{\lbrace f_{g_1},f_{g_2}\rbrace }$ denote the values of $p_{g_1}$ and $p_{g_2}$  by solving $f_{g_1},f_{g_2}$ from equations \eqref{subsection_eq_5}-\eqref{subsection_eq_6}.  
Next, we explain how to find the corners of power regions, which satisfy all the constraints. The approximate solution lies in one of the following seven regions: 

Region 1: When $p_c^{*}= P_c^{\max}$,
solve equations \eqref{subsection_eq_5}-\eqref{subsection_eq_7} by putting value of $p_c^{*}= P_c^{\max}$, values of $p_{g_1},p_{g_2}$  lies on any of one of these corners. 
The value of $p_{g_1}$ and $p_{g_2}$ is found by tracing these points
\begin{equation}
p_{g_1},p_{g_2} \in \Big(\lbrace p_{g_1},p_{g_2}\rbrace_{\lbrace f_{g_1},f_{g_2}\rbrace }, P_c^{\max}\Big) 
\bigcup \Big(\lbrace p_{g_1},p_{g_2}\rbrace _{\lbrace f_{g_1},f_{c}\rbrace }, P_c^{\max}\Big)\bigcup \Big(\lbrace p_{g_1},p_{g_2}\rbrace _{\lbrace f_{c},f_{g_2}\rbrace }, P_c^{\max}\Big) \nonumber 
\end{equation}

Region 2: When $p_{g_1} = P_g^{\max}$, solve equations \eqref{subsection_eq_5}-\eqref{subsection_eq_7} by putting value of $p_{g_1}^{*}= P_{g_1}^{\max}$, values of $p_c, p_{g_2}$ lies on any of one of these corners, i.e,
\begin{equation}
\begin{aligned}
p_{c},p_{g_2} &\in \left(\lbrace p_{c},p_{g_2}\rbrace _{\lbrace f_{g_1},f_{g_2}\rbrace }, P_{g_1}^{max}\right) 
\bigcup \left(\lbrace p_{c},p_{g_2}\rbrace _{\lbrace f_{g_1},f_{c}\rbrace }, P_{g_1}^{max}\right)
\bigcup \left(\lbrace p_{c}, p_{g_2}\rbrace _{\lbrace f_{c},f_{g_2}\rbrace }, P_{g_1}^{max} \right) \nonumber
\end{aligned}
\end{equation}

Region 3: When $p_{g_2} = P_g^{\max}$, solve equations \eqref{subsection_eq_5}-\eqref{subsection_eq_7} by putting value of $p_{g_2}^{*}= P_{g_2}^{\max}$, values of $p_c, p_{g_1}$ lies on any of one of these corners, i.e,
\begin{equation}
\begin{aligned}
p_{c},p_{g_1} &\in \left(\lbrace p_{c},p_{g_1}\rbrace _{\lbrace f_{g_1},f_{g_2}\rbrace }, P_{g_2}^{max}\right)
\bigcup \left(\lbrace p_{c},p_{g_1}\rbrace _{\lbrace f_{g_1},f_{c}\rbrace }, P_{g_2}^{max}\right)
\bigcup \left(\lbrace p_{c}, p_{g_1}\rbrace _{\lbrace f_{c},f_{g_2}\rbrace }, P_{g_2}^{max} \right) \nonumber
\end{aligned}
\end{equation}
Also, there exist other solutions, that is, when two of three co-channel users transmit at the maximum power and constraint of required SINR threshold are fulfilled.

Region 4: When  $p_c^{*}= P_c^{\max}$ and $p_{g_1} = P_g^{\max}$, then from \eqref{subsection_eq_6}, $p_{g_2} = \left(P_{g_2,\lbrace f_{g_2}\rbrace},P_{g_1}^{max}, P_{c}^{max} \right)$ 

Region 5: When  $p_c^{*}= P_c^{\max}$ and $p_{g_2} = P_g^{\max}$, then from \eqref{subsection_eq_5}, $p_{g_1} = \left(P_{g_2}^{max}, p_{{g_1,\lbrace f_{g_1}\rbrace}},P_{c}^{max} \right)$

Region 6: When  $p_{g_1} = P_g^{\max}$ and $p_{g_2} = P_g^{\max}$, then from \eqref{subsection_eq_7},
$p_{c} = \left(P_{g_2}^{max}, P_{g_1}^{max}, p_{c,\lbrace f_{c}\rbrace} \right)$

There is a final region, Region 7, where all the transmitters are transmitting at the maximum power and SINR thresholds are fulfilled, that is, $p_c^{*}= P_c^{\max}$, $p_{g_1} = P_{g_1}^{\max}$ and $p_{g_2} = P_{g_2}^{\max}$.\\
By tracing these points, we numerically prove that an approximate solution with low computational complexity in comparison to the exhaustive search is achieved.

The example in Appendix \ref{appndx:examples_2} demonstrates the use of the corner search method for the power allocation for the scenario when $G_k$ = 2.

Now, we discuss the solution of the power allocation problem for any arbitrary number of MGs sharing a channel with a CU.
\subsection{Power allocation for an arbitrary number of MGs on a channel}
In this subsection, we discuss the general case where any number of MGs may share the channel with a CU. From the channel allocation algorithms, we know the number and identities of the MGs that share any channel. Since, we are managing the interference between two simultaneously operating technologies (cellular and D2D), we call the proposed scheme as Simultaneous Transmission Interference Management (STIM) scheme.
\subsection*{Simultaneous Transmission Interference Management (STIM) Scheme}
\label{Sec:6_wcnc_17} 
We derive the power transmitted by the MGs using modified ``Generalized Distributed Constrained Power Control Algorithm (GDCPC)'' \cite{im2008autonomous}, which iterates to fulfill the SINR requirement of every user while limiting the interference to other receiving nodes upto a tolerable level. Initially, each MGTX calculates its transmit power by considering only its channel gain to eNB and achievable QoS by sharing channel with the corresponding CU. However, this derived power may not be optimal, as a MG may not know the transmit power of other MGs which are sharing the same channel and causing interference. If we distribute the total interference to a CU into individual constraints of MGs, the QoS of every CU can be ensured. In other words,
\begin{equation*}
\label{power_control_cu_wcnc_17}
h_{g,b}^k p_{g,k} \leq \dfrac{I_{\textrm{th}}^k}{G_k} \quad \forall g \in G_{k},
\end{equation*}
where $I_\textrm{th}^k = \sum_{g=1}^{\vert \mathcal{G}_k\vert} p_{g,k} h_{g,b}^k $. The maximum power that can be allocated to a MG satisfying the shared CU QoS constraint can be calculated as:
\begin{equation}
\label{P_g_wcnc_17}
p_{g,k} =  \dfrac{I_{th}^{k}}{h_{g,b}^{k} G_{k}}
\end{equation} 
From \eqref{P_g_wcnc_17}, it can be observed that the
BS should know the other MGs which are sharing this same channel, channel gain from MG to the BS, and the maximum interference threshold of the co-channel CU. 
This information can be obtained from the channel allocation algorithm above. In addition, the BS periodically transmits beacon signals, thus with channel reversal, the BS knows the channel gain from MGTX to the BS. The maximum transmit power 
of the $g^\textrm{th}$ MG can be expressed as 
\begin{equation*}
\label{max_power_wcnc_17}
p_{g,k}^{\max}= min \lbrace P_{g}^{\max}, p_{g,k} \rbrace
\end{equation*}
To fulfill the SINR constraints of every CU and MG, the transmission power on the allotted channel is updated as follows:
\begin{equation}
\label{updation}
 p_{g,k}{\left(t\right)} = \left \{
  \begin{aligned}
    &\dfrac{\Gamma_{g}^{th}}{\Gamma_{g,k}\left(t-1\right)} , && \text{if} \dfrac{\Gamma_{g}^{th}}{\Gamma_{g,k}\left(t-1\right)}p_{g,k}\left(t-1\right) \leq P_{g}^{\max} \\
    &{p^*_{g,k}} , && \text{otherwise}
  \end{aligned} \right.
\end{equation}
where ${p^*_{g,k}}$ can be calculated as follows:
\begin{equation}
\begin{aligned}
\label{arbitrarily_chosen}
{p^*_{g,k}} &= min \lbrace \overbrace{p_{g,k}}, p_{g}^{\max} \rbrace, \end{aligned}
\end{equation}  
where $\overbrace{p_{g,k}}$ lies within $\left[0, {p^{*}_{g,k}}\right]$. 

By knowing the channel allocation, the BS can broadcast the maximum power allocated on every channel, and  then every mobile device updates its transmission power using \eqref{updation}. The whole process of channel allocation using either criterion and power allocation to D2D MGs allotted to different channels is summed-up in Algorithm \ref{algo_2_wcnc_17}. 

\begin{algorithm}[t]
  \caption{Interference-aware and outage-aware channel allocation and STIM based power allocation}
  \label{algo_2_wcnc_17}
 \DontPrintSemicolon
\Begin($ $)
{
For every MGTX $g \in \mathcal{G}_k$, find the channel quality from previous time slot and determine $h_{g,b}^k \forall g \in \mathcal{G}$ and $k \in \mathcal{K}$ \;

\textbf{Initialize} t=0, $p_{g,k} = \dfrac{P_{g,k}^{\max}}{G_k} \forall	 g \in \mathcal{G}_k$\;

\For{k = 1 to $C$}{
sort the $\Delta R_{g,k}$ in descending order}
\For{$g \in \mathcal{G}$}{
 update t = t + 1 \;
 Obtain the value of $G_k$ which are going to use the same channel with a CU using Algorithm~\ref{algo_wcnc_17}\;
Update the transmission power using equation \eqref{updation} 
\If{$\Delta R_{g,k} > 0$}
{
 update set $\mathcal{G}^{'}_k = \mathcal{G}^{'}_k \cup g$  
}
}
\textbf{For interference-aware channel allocation}\\
\For{$\forall k \in \mathcal{C}$}
{
j = $\mathop{\text{argmax}}_{j \in \mathcal{G}^{'}_k} \Delta R_{g,k}$\;
$\mathcal{G}^{'}_k =  \mathcal{G}^{'}_k \setminus j$\;
$\mathcal{V}_1 = \dfrac{h_{j,r}^{k,g}}{h_{j',j}^{k}}$, $\mathcal{V}_2 = \dfrac{h_{j',r'}^{k}}{h_{j,j'}^{k}}$ $\forall j' \in  \mathcal{G}^{'}_i $;

\If{$\mathcal{V}_1 > \gamma_{th}$ and $\mathcal{V}_2 > \gamma_{th}$}{
$\mathcal{G}_k = \mathcal{G}_i \cup j$, and $\mathcal{G}_k = \mathcal{G}_k \cup k$
}
}
\textbf{For outage-aware channel allocation}
\For {k = 1 to $C$} 
{
\For {j = 1 to  $\mathcal{G}^{'}_k$}
{ 
i = $\mathop{\text{argmin}}_{k \in \mathcal{C}}$ 
\{\textrm{Objective 1 or Objective 2 or Objective 3}\} \\
 $\mathcal{G}^{'}_k =  \mathcal{G}^{'}_k \setminus k$ \\
$\mathcal{V}_1 = \sum_{g=1}^{\mathcal{G}_k} p_{g,k} h_{g,b}^k$\\

\If {$\mathcal{V}_1 \leq I_{th}^k$}
{
$\mathcal{G}_k = \mathcal{G}_k \cup j$, and $\mathcal{G}_k = \mathcal{G}_k \cup k$}
}}
}
\end{algorithm}

\section{Performance Evaluation}
\label{Sec:7}
In this section, we first analyze the numerical results obtained from solving our joint channel and power allocation schemes for special cases: when a MG shares channel with exactly one CU, and when two MGs share channel with one CU. Then, we discuss the performance of the proposed schemes for general problem, where more than two MGs share channel with one CU.

\textit{General Settings:} In simulations, we consider a circular cell of radius $R$ with uniformly distributed MGs and CUs, and eNB in the center. The path loss model between CU and eNB, and among D2D MGs is considered as per 3GPP standard \cite{kuo2015new}. Unless otherwise specified, the main simulation parameters are set as those given in Table \rm{2}. We assume that all CUs have identical data rate demands and the total bandwidth is equally shared by all channels. The results are averaged over 500 Monte-Carlo simulations.

\renewcommand{\arraystretch}{0.9}
\begin{table}[!t]
\setlength\extrarowheight{2.5pt}
\caption{Simulation Parameter} 
\centering
\begin{tabular}[width=\linewidth]{|p{2.50in}|p{1.25in}|}
\hline
Cell radius (R) & 500,800 m\\
\hline
No. of cellular users  $\vert \mathcal{C} \vert$ & 3\\
\hline
Path loss exponent & 3.6 \\
\hline
Noise power & -114 dBm \\
\hline
Shadowing Std. Dev. & 8 dB \\
\hline 
Bandwidth($B_i$) & $10^6$ Hz\\ 
\hline
$P_c^{\max}$,$P_g^{\max}$ & 30 dBm \\
\hline
\end{tabular}
\label{Table_2}
\end{table}
In order to evaluate the improvements offered by the proposed algorithms,  we compare their performance with three other schemes. Authors in \cite{Dfeng} proposed two schemes. The first scheme is basically a bipartite graph based optimal matching scheme. If number of MGs is larger than the number of CUs, this scheme chooses $C$ D2D MGs that maximize the achievable system throughput by utilizing Hungarian algorithm \cite{kuhn2005hungarian}. We call this scheme ``Bipartite graph allocation''. The second scheme randomly selects $C$ MGs to reuse all subchannels and they all transmit at the maximum power. We call this scheme ``Random channel allocation''. The third scheme for comparison is Greedy heuristic scheme \cite{zulhasnine2010efficient}, where the eNB pairs-up CUs and D2D MGs whose channels have the lowest CU-MGRX interference, and all mobile nodes transmit at the maximum power. 

\subsection{A single MG shares channel with exactly one CU, $G_k$=1}
\begin{figure}[t]
\centering
\includegraphics[width=3.5in]{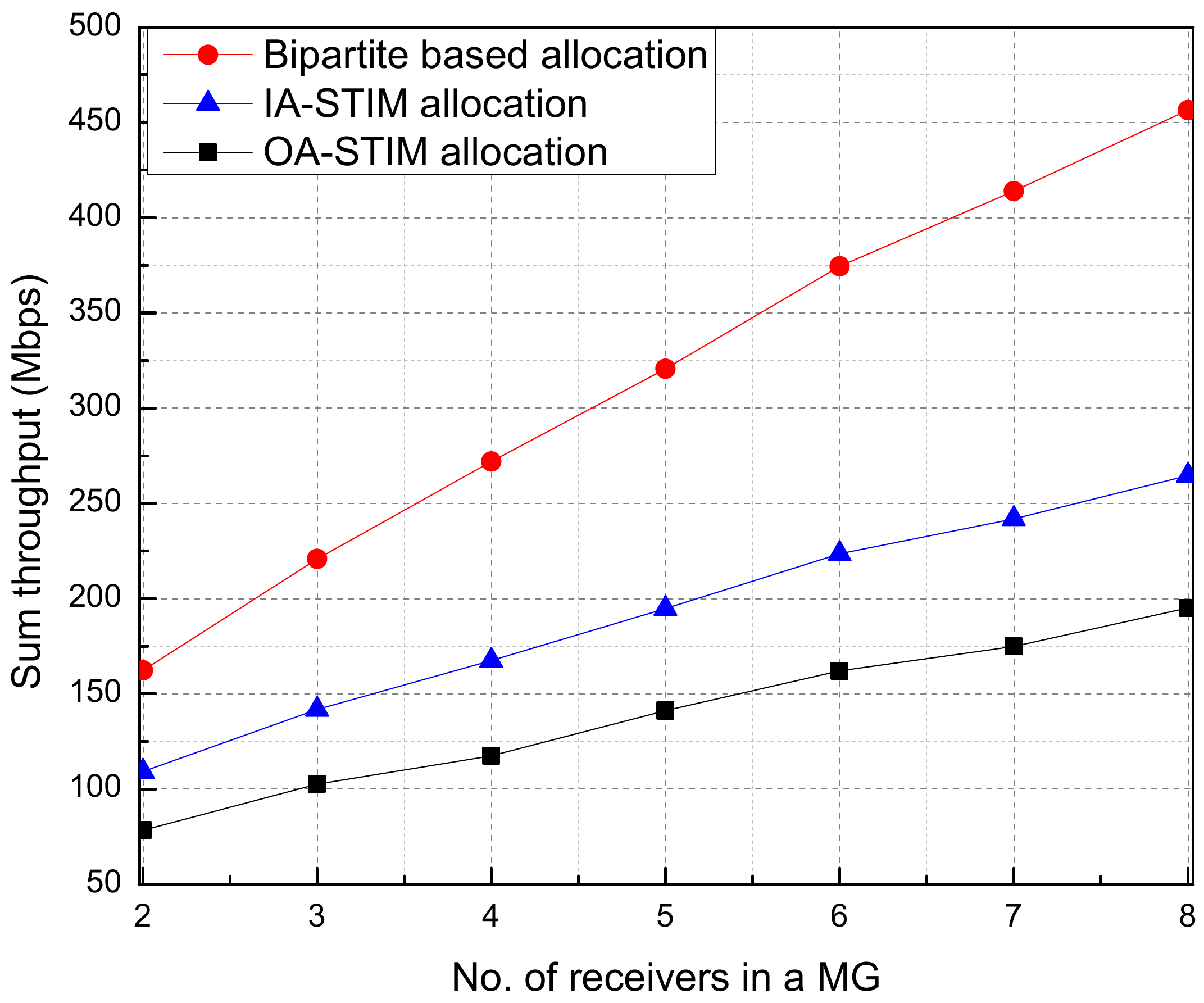}
\caption{The sum throughput variation with number of receivers in a MG, when exactly one MG share the channel with a CU, $\mathcal{C}=5,\mathcal{G} = 5$, Bi=1 MHz}
\label{fig:sum_th_num_receivers}
\end{figure}
Figure \ref{fig:sum_th_num_receivers} depicts the sum throughput variation with number of receivers in a MGs, when exactly one MG shares channel with a CU. It can be observed that the sum throughput increases sublinearly with increase in number of receivers in a MG. The reason for this is as the number of receivers in each MG increases, the worst receiver in each MG may contribute poorly to the sum throughput, and that determines the throughput of the MG. It can also be observed that the sum 
throughput value is higher for bipartite based power allocation in comparison to IA-STIM and OA-STIM. This is because when number of MGs are less than or equal to number of CUs, the bipartite based matching is optimal, as either of the CU or MG transmits at the maximum power, as proved in Lemma \ref{ch_2_lemma_1}. While in case of IA-STIM, to maintain the interference threshold below certain value, the maximum transmit power of MGTX is always low, and in case of OA-STIM, to maintain the outage threshold of CUs and to support the stringent QoS requirement of every receiver in the MG, MGTXs transmit at low power. 

\subsection{Two MGs share the channel with one CU, $G_k=2$}
\begin{figure}
\centering
\includegraphics[width=3.5in]{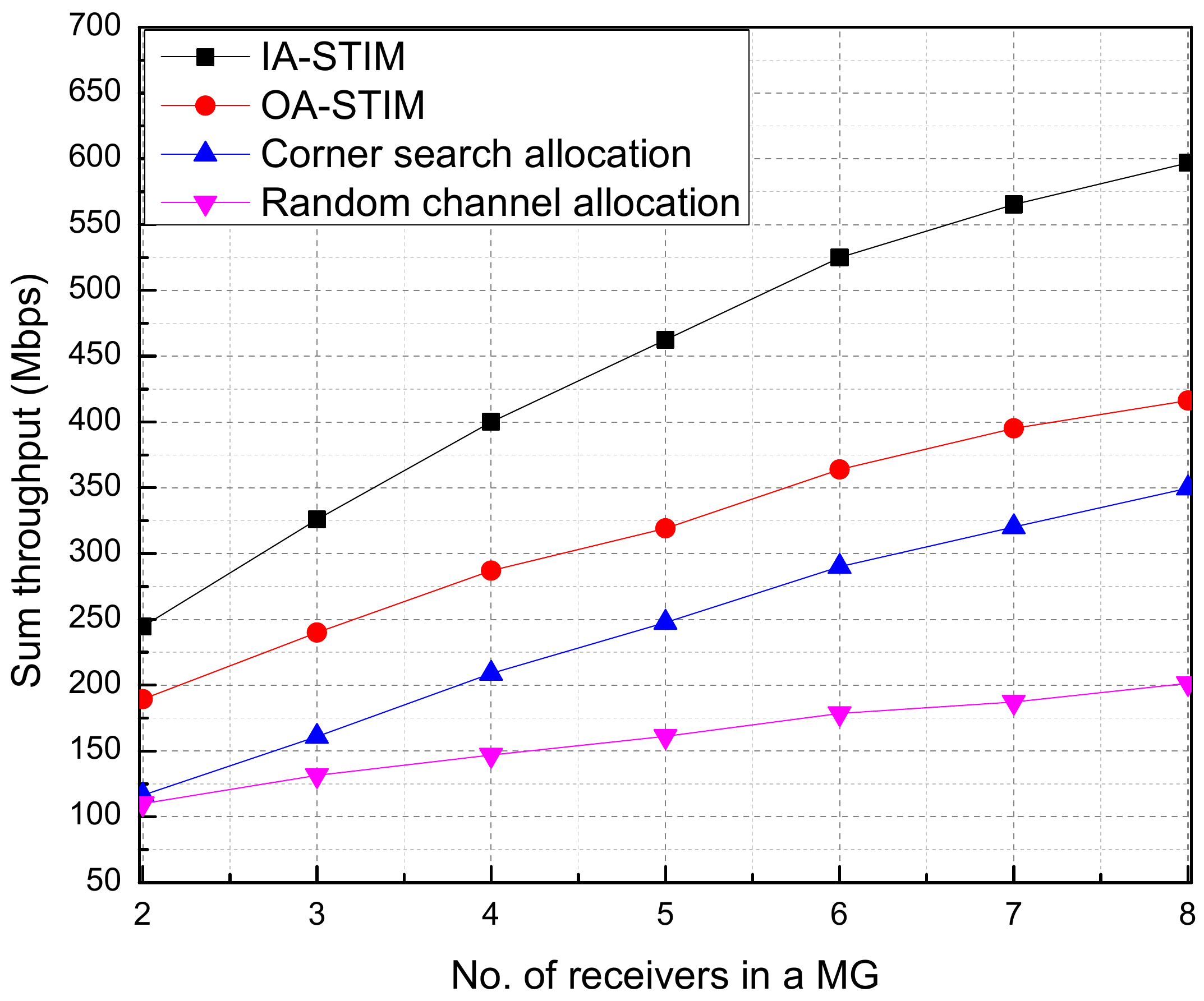}
\caption{The sum throughput variation with number of receivers in a MG, when exactly two multicast groups share resources with a cellular user, $\mathcal{C} = 5$ $Bi$ = $10^6$ Hz.}
\label{fig:gk_2_receivers}
\end{figure}

Figure \ref{fig:gk_2_receivers} depicts the sum throughput variation 
with number of receivers in a MG for scenarios where exactly two MGs share a channel with a CU. 
Note that in this case, random channel allocation scheme also allows two MGs to share a channel with a CU. 
It can be observed that the sum throughout increases sublinearly with increase in number of receivers in a MG, then starts saturating. The reason for this is as number of receivers in a MG increases, the worst receiver in the MG contributes poorly to the sum throughput, and that determines the throughput of the MG. In addition, it can be observed that the performance of IA-STIM is better than that of OA-STIM, Corner search, and Random channel allocation schemes. The difference between performance of IA-STIM and OT-STIM is because of interference handling algorithm. In IA-STIM, only those groups are paired which have minimum mutual interference along with maintaining SINR thresholds of the CUs, while in OA-STIM, those MGs are paired which minimize the outage thresholds. Therefore, IA-STIM allows more power allocation to MGs which leads to higher SINR, and concomitantly higher contribution to the sum throughput. The performance of corner search algorithm is poorer with respect IA-STIM and OA-STIM because it tries to support minimum SINR thresholds of CUs and MGs, and searches the optimal solution only among limited feasible solutions, although it is computationally efficient and thus, may serve as a good heuristic to efficiently provide a lower bound to the optimal throughput.

\begin{figure}
\centering
\includegraphics[width=3.5in]{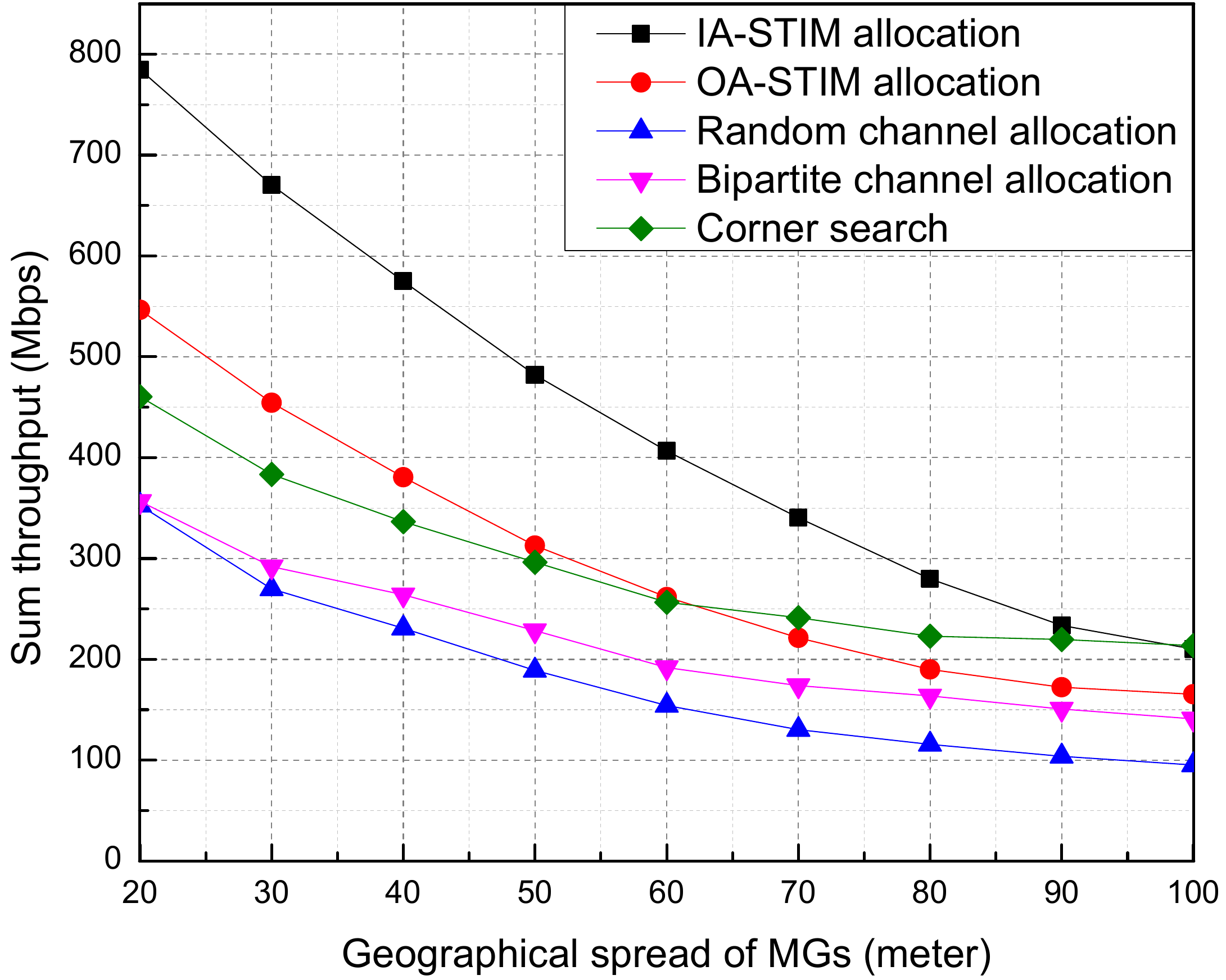}
\caption{The sum throughput variation with geographical spread when when exactly two multicast groups share resources with a cellular user.}
\label{fig:4_corner}
\end{figure}

Figure \ref{fig:4_corner} depicts the sum throughput variation with geographical spread of MGs for different power allocation schemes. It can be observed that the performance of the proposed corner search method is close to IA-STIM, while it is highly computationally efficient.  
It can also be observed that the difference in sum throughput of IA-STIM and corner search decreases with increase in geographical spread. This is because for larger geographical spread, MGs try to allocate the maximum power in both the schemes. For lower geographical spread, the proposed OA-STIM scheme is slightly better than corner search method. However for larger spread, the corner search is better, as in case of OA-STIM, we always need to maintain the outage thresholds of primary cellular users, therefore, the MGTXs do not transmit at the maximum power. Furthermore, the performances of the proposed scheme: IA-STIM, OA-STIM and Corner search, are better than those of random channel allocation and bipartite-graph based allocation schemes.

\subsection{Many MGs shares channel with one CU, $G_k>2$}
\begin{figure}[t]
\centering
\includegraphics[width=3.5in]{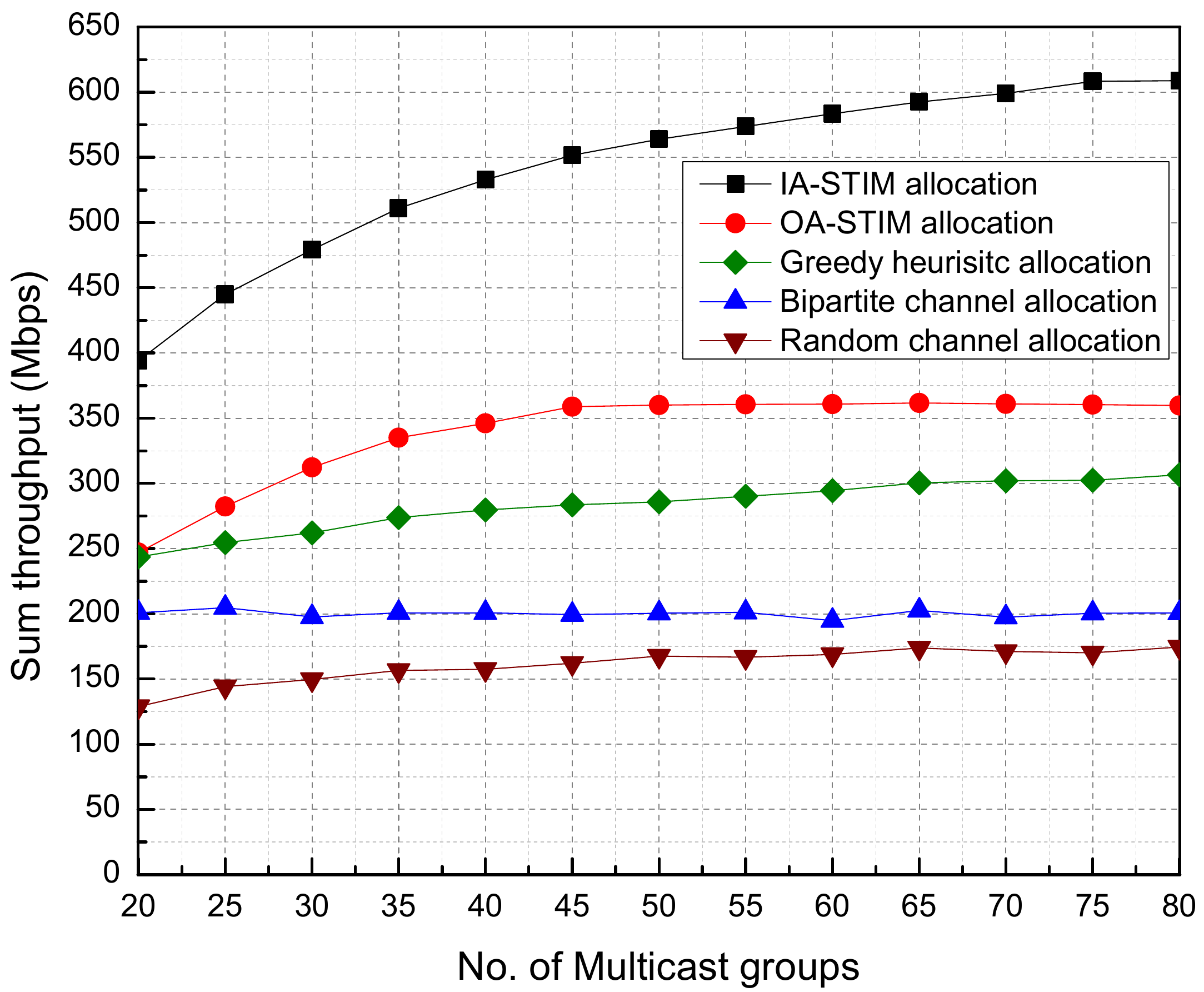}
\caption{The sum throughput with varying no. of D2D multicast groups, C = 5.}
\label{fig:6_wcnc_17}
\end{figure}

In Figure \ref{fig:6_wcnc_17}, the sum throughput variation with variable number of MGs is plotted. It can be observed that for all the schemes, the sum throughput increases with increase in number of MGs upto a certain level, then it saturates. Indeed, this trend is expected in IA-STIM, as the number of MGs increases in a cell, more and more MGs qualify the sharing criteria (mutual interference $ \leq \gamma_{th}$), hence a channel is shared by multiple MGs. However, after a certain level, the co-channel interference among MGs starts increasing. Therefore, no more MGs can be allocated to a channel. It can be observed that the performance of OA-STIM is less than the IA-STIM. The reason for this is as in OA-STIM, we need to fulfill the QoS requirement of every receiver, therefore, there is lesser chance to find a channel that satisfies the demands of multiple MGs. 

\begin{figure}
\centering
\includegraphics[width=3.5in]{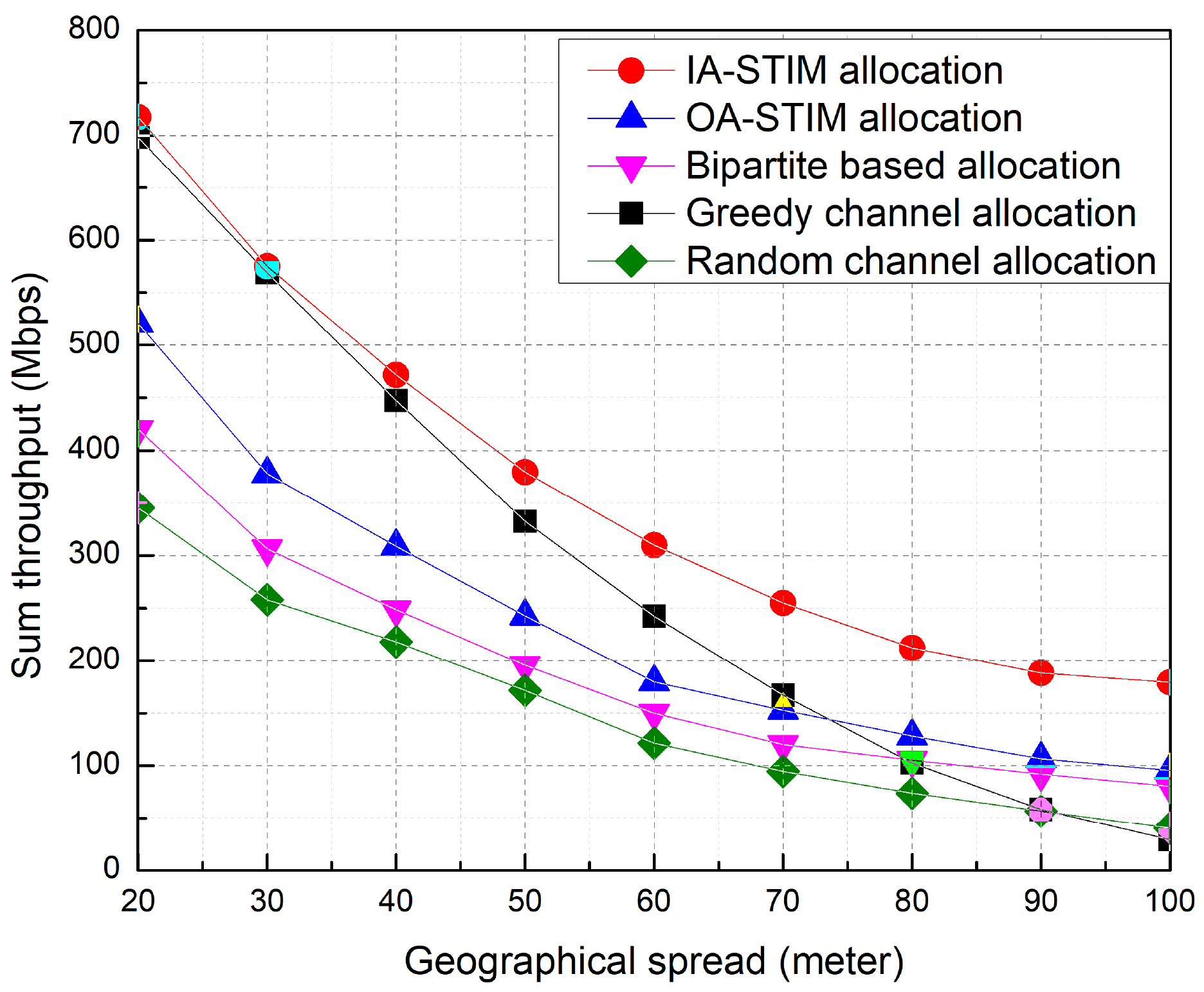}
\caption{The sum throughput with varying geographical spread of MGs, C=5, G=20.}
\label{fig:7_wcnc_17}
\end{figure}

Figure \ref{fig:7_wcnc_17} depicts the variation of the sum throughput with geographical spread of MGs for scenarios where more than two MGs share a single channel. It can be observed that the sum rate rapidly decreases with geographical spread. 
The reason for this decrease is as geographical spread increases, the strength of signal received by MG receivers decreases, and there is higher probability that there exists a receiver having very low SINR, which defines the sum rate of a MG. Furthermore, each MGTX tries to increase the transmit power for fulfilling the minimum threshold of D2D MGs. However, this may create severe interference to CUs, which is not permissible. It can be noted that for smaller geographical spread, the performances of IA-STIM and Greedy schemes are almost the same. This is due to interference avoidance mechanism in both the schemes.
Similar to the proposed IA-STIM scheme which tries to avoid MGs having high mutual interference to share a single channel, greedy scheme also tries to avoid sharing the channel which has severe interference from CU to MG receiver. Therefore, they have high value of SINR, and consequently higher sum throughput. It can also be observed that the performance of the proposed scheme is better than Bipartite graph allocation, Random channel allocation, and Greedy heuristic allocation for short-range communication.

\begin{figure}
\centering
\includegraphics[width=3.5in]{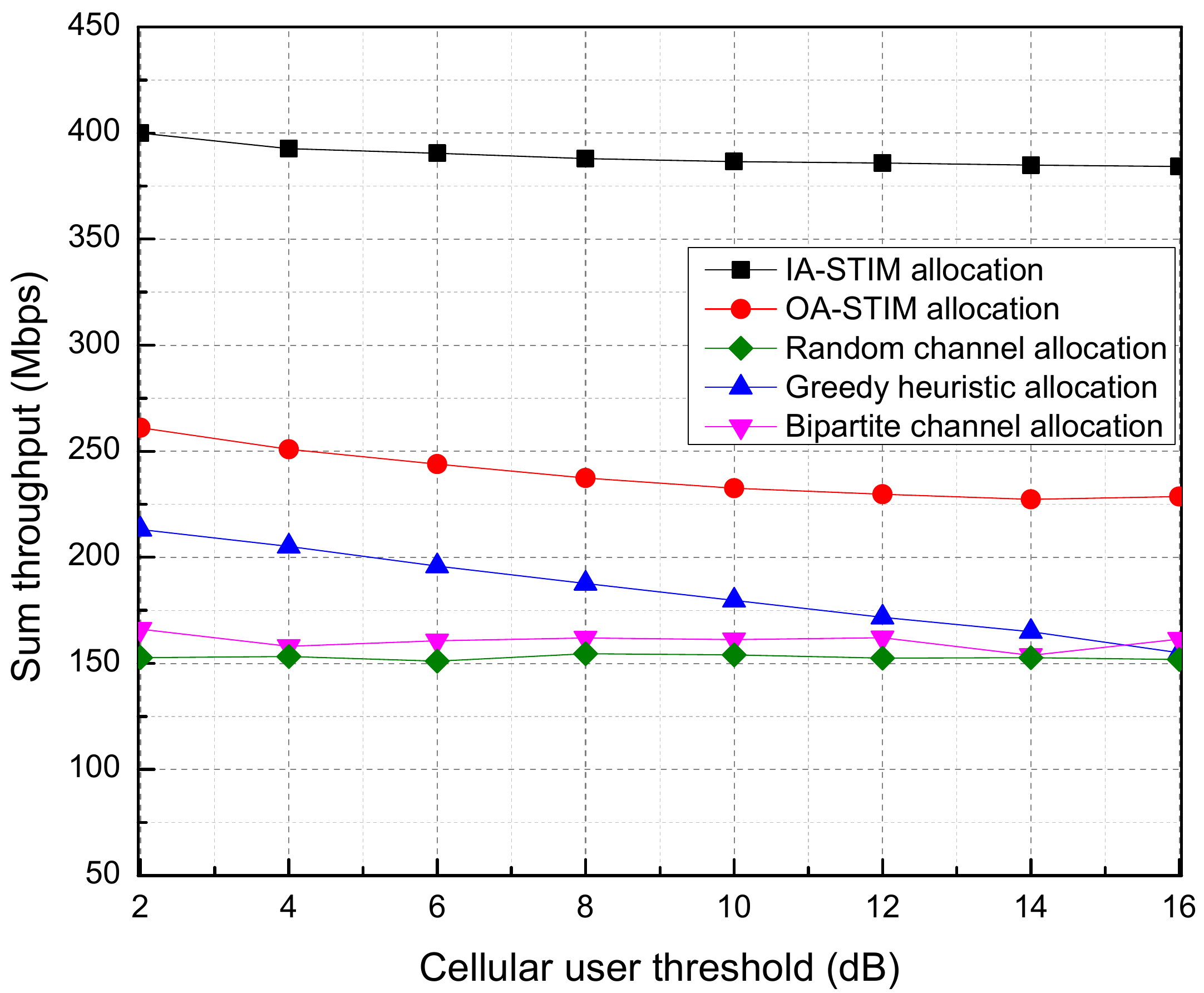}
\caption{The sum throughput as a function of each CU's QoS requirement, C=5, G=20.}
\label{fig:8_wcnc_17}
\end{figure}

Figure \ref{fig:8_wcnc_17} depicts the sum throughput as a function of CUs' QoS requirements for the case when more the two MGs share a channel. It can be observed that with increase in QoS threshold, sum throughput decreases, and it becomes almost constant in range of 12 - 16 dB. When QoS requirements of CUs are high, CUs try to transmit at the maximum power, causing significant interference to MGs, and maximum power that can be allocated to MGs starts to decrease (to avoid outage of CUs). Therefore, their contribution in the sum throughput also decreases. The performance of IA-STIM and OA-STIM is better than all other schemes. When the QoS requirement of CUs is too high, we can not share the channel with other MGs.  
\begin{figure}[t]
\centering
\includegraphics[width=3.5in]{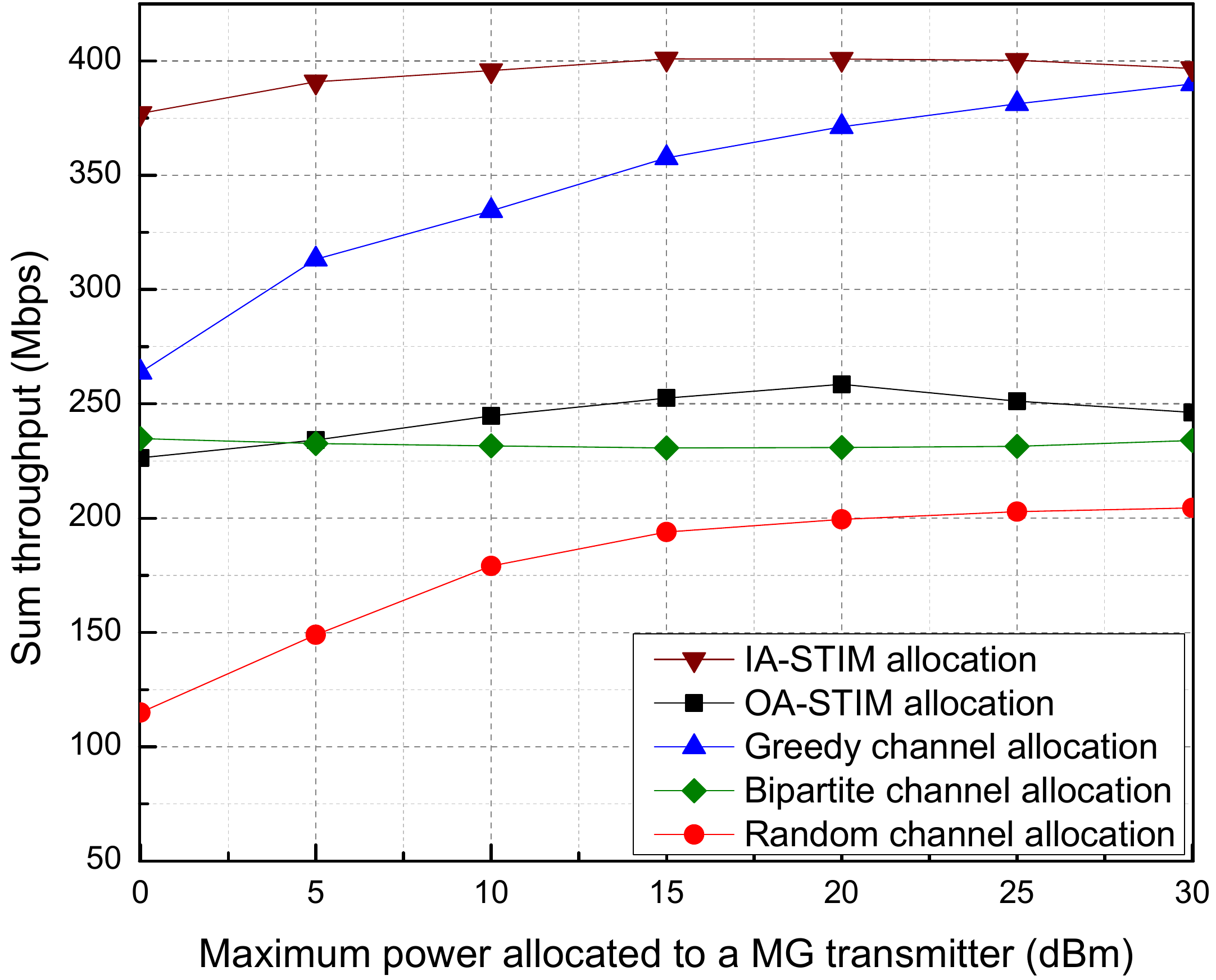}
\caption{The sum Throughput as a function of maximum transmission power, C = 5 , G=20, R=500m.}
\label{fig:9_wcnc_17}
\end{figure}

Figure \ref{fig:9_wcnc_17} depicts the sum throughput variation with the maximum power allocated to MGTXs. It can be observed that for the proposed IA-STIM and OA-STIM schemes, the sum throughput initially increases with increase in the maximum transmission power, and then it starts to decrease. The reason for this is, when the MGs transmit at low power, their interference to co-channel MG's receiver is low, thus more and more MGs can be supported per channel. However, as MGs have lower transmit power, the receivers have lower SINR too. Therefore, their contribution to the sum throughput is also low. 
With increasing transmission power, the contribution of MGs to the sum throughput increases, however, after certain power level, it start decreasing. The reason for this is, for higher transmission power, MGs create interference to the receivers of co-channel MGs, therefore, MGs' contribution to the sum throughput decreases. Specifically, after 15 dBm, interference to MGs' receivers becomes excessive, therefore, MGs' contribution to the sum throughput decreases. This observation helps in designing the systems such that maximum power per MGTX can be limited to 15 dBm for the maximum sum throughput.

\section{Conclusion and Future Work}
\label{Sec:8}
We address the uplink resource allocation problem in fully-loaded D2D multicast enabled underlay cellular network. To maximize the overall sum-throughput while guaranteeing the QoS for CU's and D2D MG receivers, we formulate an MINLP optimization problem. The problem is decomposed into two separate problems of channel and power allocation. Two channel allocation algorithms for D2D MGs and CUs are proposed and an efficient power allocation scheme is proposed. For two specific scenarios, where either only one or two MGs share channel with a CU, we provide power allocation schemes which are more efficient that the scheme proposed for the general case. Then, based on these scheme efficient joint channel and power allocation schemes: IA-STIM and OA-STIM, are proposed. The performance evaluation of the proposed schemes considers the dynamic MG scenarios and illustrates its sensitivity to different system parameters, including number of MGs sharing a channel, the QoS requirements of a CU, and the maximum transmission power available to multicast transmitters. The numerical results establish that the proposed schemes outperform existing resource allocation schemes in terms of sum throughput. Moreover, the performance of the proposed scheme has been found to improve as number of CUs increases since the proposed scheme achieves a multiuser diversity gain in selecting a cellular user channel.

In this paper, we only address the channel and power allocation problem in a single-cell scenario. An interesting extension would be to investigate the multi-cell system where each cell optimizes its individual performance. In addition, we only propose a centralized solution to channel and power allocation. The design and analysis of a distributed channel and power allocation schemes in this scenario may lead to more practical solutions.

\appendices
\setcounter{equation}{0}
\renewcommand{\theequation}{\thesection.\arabic{equation}}
\section{Proof of Proposition~\ref{proposition:1}}
\label{appendix_proposition}
In channel sharing and interference limited system model, the achievable rate of a CU that communicates on the $k^{\textrm{th}}$ channel with transmit power $p_{c,k}$ can be expressed as
\begin{equation}
\label{prop_0}
R_{k,g}^c{\left(k,p_{c,k}\right)} = B_k \log_2 \left( \dfrac{p_{c,k} h_{c,b}^k}{ \sum_{g=1}^{\vert \mathcal{G}_k \vert} p_{g,k} h_{g,b}^k}\right)
\end{equation}
By assuming $B_k$ = 1 and rearranging the equation \eqref{prop_0}, we get
\begin{equation}
\label{prop_1}
R_{k,g}^c{\left(k,p_{c,k}\right)} = \log_2 \left( {p_{c,k} h_{c,b}^k}\right) - \log_2 \Bigg({  \sum_{g=1}^{\vert \mathcal{G}_k \vert} p_{g,k} h_{g,b}^k}\Bigg)
\end{equation}
For $p_{g,k} = P_{g}^{\max}$, $ \log_2 \Big({ \sum_{g=1}^{\vert \mathcal{G}_k\vert}  P_{g}^{\max} h_{g,b}^k}\Big) > \log_2 \Big({ \sum_{g=1}^{\vert 
\mathcal{G}_k \vert} p_{g,k} h_{g,b}^k}\Big)$

Therefore, \eqref{prop_1} can be written as  
\begin{equation}
\label{prop_2}
R_{k,g}^c{\left(k,p_{c,k}\right)} > \log_2 \left( {p_{c,k} h_{c,b}^k}\right) - \log_2 \Bigg({  \sum_{g=1}^{\vert \mathcal{G}_k \vert} P_g^{\max} h_{g,b}^k}\Bigg)
\end{equation}
By taking summation over all channels, \eqref{prop_2} can be written as 
\begin{equation}
\sum_{k=1}^{\vert \mathcal{K} \vert} R_{k,g}^c{\left(k,p_{c,k}\right)} > \sum_{k=1}^{\vert \mathcal{K} \vert}  \log_2 \left( {p_{c,k} h_{c,b}^k}\right) - \sum_{k=1}^{\vert \mathcal{K} \vert} \sum_{g=1}^{\vert \mathcal{G}_k \vert} \log_2 \Big({  P_g^{\max} h_{g,b}^k}\Big)
\end{equation}

\section{Proof of Lemma \ref{lemma_op}}
\label{proof_OP}
\setcounter{equation}{0}
The general form of outage probability of a typical receiver of a D2D MG is 
\begin{align}
\text{Pr} \left(\Gamma_{r,g}^{k} \leq \gamma_\textrm{th}^{o} \right) &= 1 - \text{Pr} \left(h_{ij} \geq \gamma_\textrm{th}^{o} d_{g,r}^\alpha \left(I_1 + I_2 \right)\right) \nonumber \\
&=\int_{0}^{\infty} e^{-s \gamma_{th}^{o} d_{g,r}^\alpha d \left[ \text{Pr} \left(I_1 + I_2 \leq s \right) \right]} \nonumber\\
&= \psi_{I_1}\left(\gamma_{th}^{o} d_{g,r}^\alpha\right) \psi_{I_2}\left( \gamma_\textrm{th}^{o} d_{g,r}^\alpha\right) \label{psi_0}, 
\end{align}
where $I_1$ and $I_2$ denote the interference created by the CUs and co-channel MGs, with corresponding values $I_1 = p_{c,k} h_{c,r}^k$ and $I_2 =  \sum_{j=1,j\neq g}^{\vert \mathcal{G}_k \vert} p_{j,k} h_{j,r}^{k,g}$. Laplace transformation of $I_1$ and $I_2$ is denoted by $\psi_{I_1}\left(.\right)$ and $\psi_{I_2}\left(.\right)$, respectively. As $h_{ij}$ 
follows the independent exponential distribution, therefore, from \cite[Definition 4.7]{haenggi2012stochastic},
\begin{align}
\psi_{I_1} \left(s\right) = e^{-\lambda_1 \pi \left(\frac{sp_{c,k}}{p_{g,k}}\right)^{\frac{2}{\alpha}} \Gamma \left(1 + \frac{2}{\alpha}\right)\Gamma \left(1 - \frac{2}{\alpha}\right)}, \label{psi_1}\\
\psi_{I_2} \left(s\right) = e^{-\lambda_2 \pi \left(\frac{sp_{g,k}}{p_{c,k}}\right)^{\frac{2}{\alpha}} \Gamma \left(1 + \frac{2}{\alpha}\right)\Gamma \left(1 - \frac{2}{\alpha}\right)} \label{psi_2},
\end{align}
where $\Gamma\left(x\right) = \int_{0}^\infty e^{-t} t^{x-1} dt$ is the complete gamma function.  
By putting \eqref{psi_1} and \eqref{psi_2} into \eqref{psi_0}, and letting $\chi_{g,k} =  \pi \Gamma \left(1 + \frac{2}{\alpha}\right)\Gamma \left(1 - \frac{2}{\alpha}\right) $, \eqref{eq_prob} is obtained.

\section{Proof of Lemma~\ref{ch_2_lemma_1}}
\label{appndx2:lemma_1}
\begin{figure}
\centering     
\subfigure{\label{fig:1_2}\includegraphics[width=40mm]{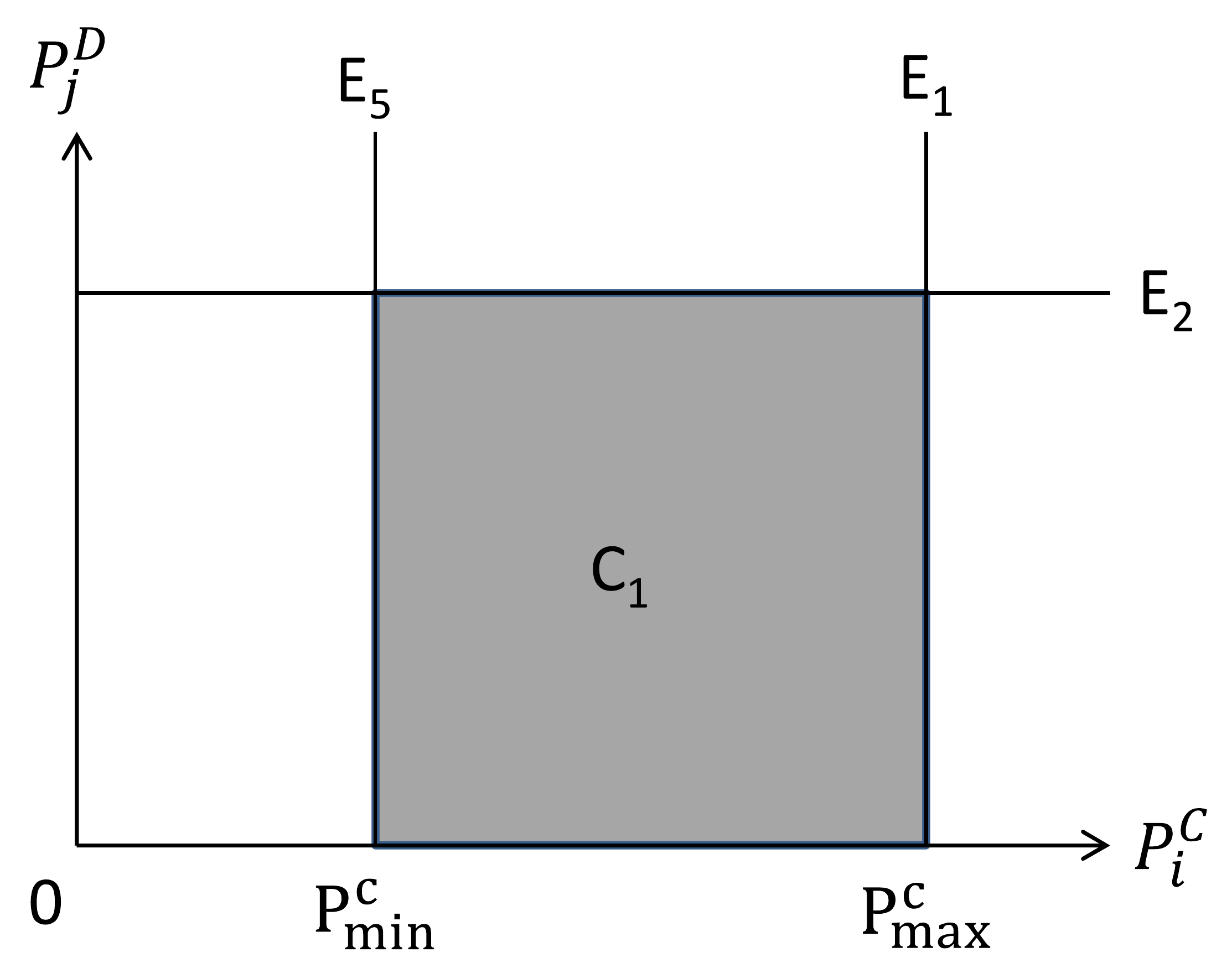}}
\subfigure{\label{fig:1_3}\includegraphics[width=40mm]{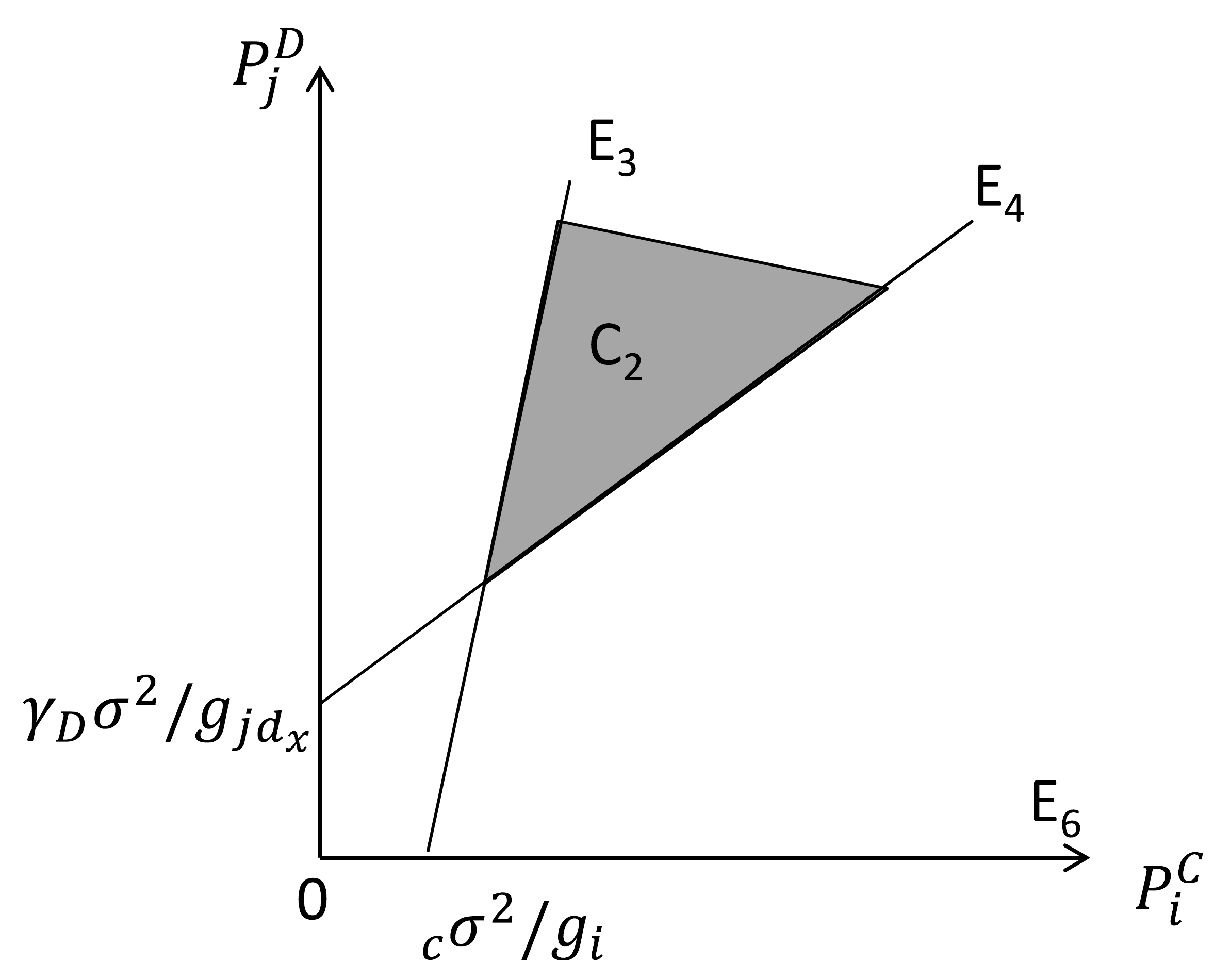}}
\subfigure{\label{fig:1_4}\includegraphics[width=40mm]{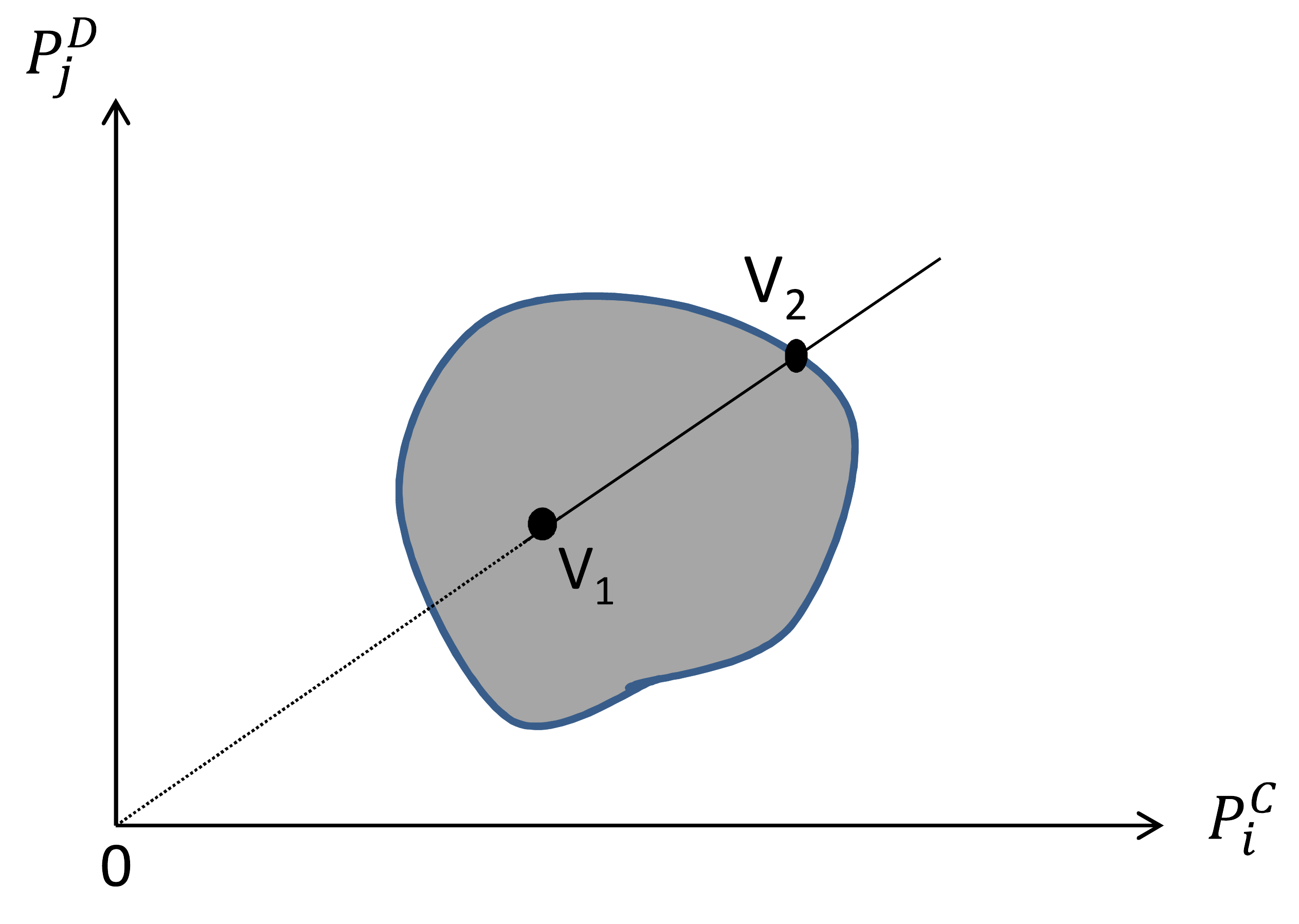}}
\caption{Feasible regions: (a) feasible region $\mathcal{C}_1$, (b) feasible region $\mathcal{C}_2$, (c) the optimal power allocation $\left(P_{i}^{*},P_{j}^{*}\right)$ achieved on the boundary of $\mathcal{W}$.}
\vspace{-0.2in}
\end{figure}

\setcounter{equation}{0}
We prove the lemma using contradiction. Consider constraints $\mathcal{C}_{1}$ and $\mathcal{C}_{2}$ in \eqref{main_problem}, as depicted in Figure \ref{fig:1_2} and Figure \ref{fig:1_3} respectively, and  $\mathcal{W} = \mathcal{C}_{1}\cap \mathcal{C}_{2}$. As $\mathcal{C}_{1}$ is finite and closed region, therefore $\mathcal{W}$ is also finite closed domain. Parameters $P_{i}^C,P_{j}^D, \Gamma_{c}, \Gamma_{D}$ are set accordingly to ensure that $\mathcal{W}$ is a non-empty set, so we assume that $\mathcal{W}$ is closed and bounded set. Let $\partial \mathcal{W}$ be the boundary region of $\mathcal{W}$, so $\mathcal{W}'$ = $\lbrace \mathcal{W}\rbrace \backslash \lbrace \partial \mathcal{W}\rbrace$. Further assuming a point $V_{1}\left(P_{i}^{*},P_{j}^{*}\right)$ in $\mathcal{W}'$  as depicted in Figure \ref{fig:1_4}, a line is drawn through point $V_{1}$ 
which intersects the boundary at $V_{2}\left(P_{i}^{b},P_{j}^{b} \right)$ with slope $m = P_{j}^{*}/P_{i}^{*}\geq 0$. Let $P_{i}^{b} = \alpha P_{i}^{*} \text{ and } \alpha > 1$. Since $m$ is $\frac{P_{i}^{b}-P_{i}^{*}}{P_{j}^{b}-P_{j}^{*}}$, so $P_{j}^{b}= \alpha P_{j}^{*}$. 
By putting value $\left(\alpha P_{i}^C,\alpha P_{j}^D\right)$ in \eqref{eq:7_ch1}, for $\alpha>1, \alpha \: \in \: {\mathcal{R}^{+}}$ and $\left(P_{i}^C,P_{j}^D\right)\in \mathcal{ W}$.

\begin{equation*}
\label{eq_1_8}
C\left(\alpha P_{i},\alpha P_{j}\right)= {\log}\left(\left(1+ \beta_{1} \right)\left(1+ \beta_{2} \right)^{X}\right) > C\left(P_{i},P_{j}\right),
\end{equation*}
where, $\beta_{1}= 
\frac{P_{i}^C h_{c_i}}{\sigma^{2}/\alpha+P_{j}^D h_{j}}$ 
and $\beta_{2} = \frac{P_{j}^D h_{jd_{x}}}{\sigma^{2}/\alpha+P_{i}^C h_{id_{x}}}$.

With increasing $\alpha$, $C\left(\alpha P_{i}^{*},\alpha P_{j}^{*}\right) > C\left(P_{i}^{*},P_{j}^{*}\right)$
which contradicts the assumption that $\left(P_{i}^{*},P_{j}^{*}\right)$ is the optimal solution, thus $\left(P_{i}^{*},P_{j}^{*}\right) \in \: \partial \mathcal{W} $ holds.

\section{Proof of Lemma~\ref{ch_2_lemma_2}}
\label{appndx2:lemma_2}
\begin{figure}
\centering
\includegraphics[width=3.5in]{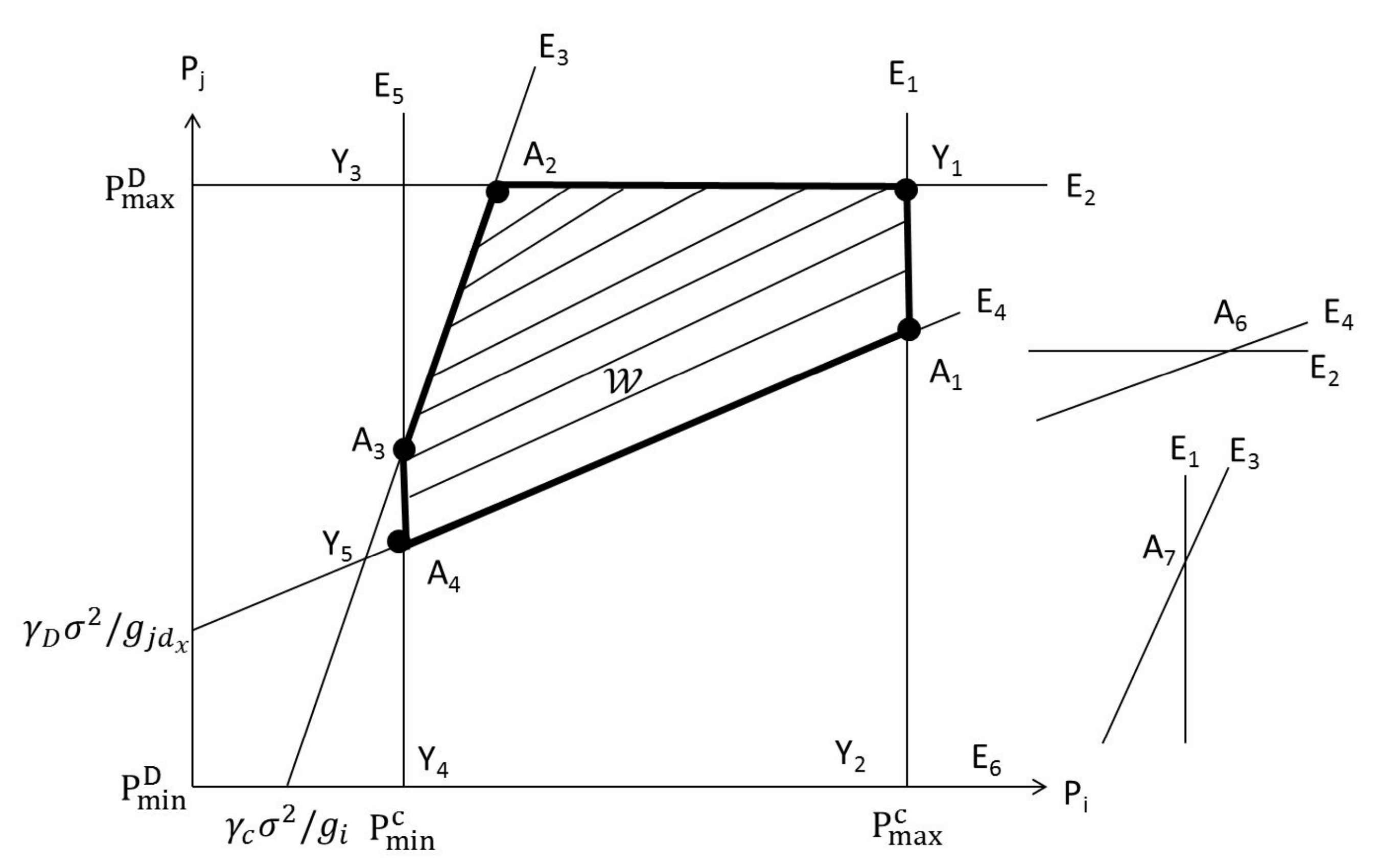}
\caption{Feasible region $\mathcal{W}$ for optimal power allocation$\left(P_{i}^{*},P_{j}^{*}\right)$}
\label{feasible_region3}
\end{figure}

\setcounter{equation}{0}
Let $\partial\mathcal{W}$ be the boundary of region $\mathcal{W}$. The contours of $\mathcal{ W}$ and $\partial \mathcal{ W}$ change with the values of constrained parameters. However, according to the problem formulation, $ \partial\mathcal{W} $ is bounded by these six edges as depicted in Fig. \ref{feasible_region3}.

Edges are $E_{1} \colon P_{i}=P_{max}^c,E_{2} \colon P_{j}=P_{max}^D,E_{3} \colon \Gamma_{i}=\gamma_{th}^{c},E_{4} \colon \Gamma_{j}=\gamma_{th}^{D},E_{5} \colon P_{i}=P_{min}^{c},E_{6} \colon P_{j}=P_{min}^{D}$.
Let $E_{n}^{'}=E_{n}\cap\partial \mathcal{W} \left(n= 1\text{ to }6\right)$ and $\mathcal{T} \left(P_{i},P_{j}\right)=\left(1+\Gamma_{i}\right)*\left(1+\Gamma_{j}\right)^{X}$. If $p_{i},p_{j} \in E_{2}^{'}$ then $\dfrac{\partial^{2}\mathcal{T}}{\partial P_{i}^{2}}\geq 0.$ $\left(P_{i},P_{j}\right) \in E_{1}^{'}\cup E_{5}^{'}, \dfrac{\partial^{2}\mathcal{T}}{\partial P_{j}^{2}}\geq 0$. If $ \left(P_{i},P_{j}\right) \in  E_{3}^{'} \cup E_{4}^{'}$, then $\mathcal{T}$ is always an increasing function. By proving $\mathcal{T}$ is a convex function, and logarithm is a monotonically increasing function, we can conclude that $\left(P_i, P_j\right)$ only resides on the corners of $\mathcal{W}$.

\section{Example for Maximum Weighted Bipartite Matching algorithm}
\label{appndx:examples_1}
\setcounter{equation}{0}
Consider a network with $C=5$ and $G=5$. 
According to Problem P2, we need to allocate the channels to the MGs with an objective of sum throughput maximization. 
First, the equality subgraph $G_(a,b)$ for a weighted cover (a,b) is the subgraph of $R_{k,g}$ whose edges are the pairs $a_i,b_j$ such that $a_i + b_j - R_{i,j} = 0$. In the cover, the excess for i and j is $a_i + b_j - R_{i,j}$, this helps in drawing equality subgraph, and covering all the vertices with least number of edges. 
X denotes the number of channels (rows) and Y denotes the number of MGs (columns). Let $Q =  T \cup R$ be a minimum vertex cover of a graph G, where $R = X \cap Q$ and $T = Y \cap Q$.

\begin{algorithm}
 \caption{Maximum Weighted Bipartite Matching}
 \label{algo_bipartite}
\DontPrintSemicolon
\KwIn{$R_{k,g}$}
\Begin($ $)
{
Represent the X and Y as the bipartition sets.\;
Initialize two variables $a_i = \max \{ R_{i,j}: i = 1,\ldots, C\}$, and $b_j=0$, j={1,2,...,G}\;
Let $Q =  T \cup R$ be a minimum vertex cover of graph G, where
$R = X \cap Q$ and $T = Y \cap Q$.\;
Derive the elements of excess matrix ($e_{i,j}$) by using this expression, $ \quad e_{j,i} =  a_i + b_j - R_{i,j}.$\;
Create a equality subgraph $G_{a,b}$ with nodes i and j which having ($e_{i,j} = 0$). Select the edge having maximum edge weight in graph G and optimal matching $O_{\max}$. If it covers all vertices of the graph, this is optimal matching\;
\If{$O_{\max}$ is \textrm{optimal matching}}
{
Stop and $O_{\max}$ is optimal matching, ($a,b$) as minimum cost cover
}
\Else
{
Calculate the step size $ \delta = \min \left(a_i + b_j - R_{i,j}\right) \{ x_i \in X \backslash R , y_i \in Y \backslash T \}$. 
Update $a$ and $b$:
$a_i :=  a_i - \delta$ if $x_i \in X \backslash R$\\
$b_j := b_i + \delta$ if $ y_j \in T$
}
}
\end{algorithm}

Let the $R_{k,g}$ matrix of size $C \times G$ is as follows:

\begin{equation*}
R_{g,k} = \begin{pmatrix} 1 & 2 & 3 &4 &5\\ 6 & 7& 8 & 7& 2
\\ 1 & 3& 4 & 4& 5\\ 3 & 6& 2 & 8& 7
\\ 4 & 1& 3 & 5& 4 \end{pmatrix}
\end{equation*}

\begin{figure}
\centering
\includegraphics[width=3.5in]{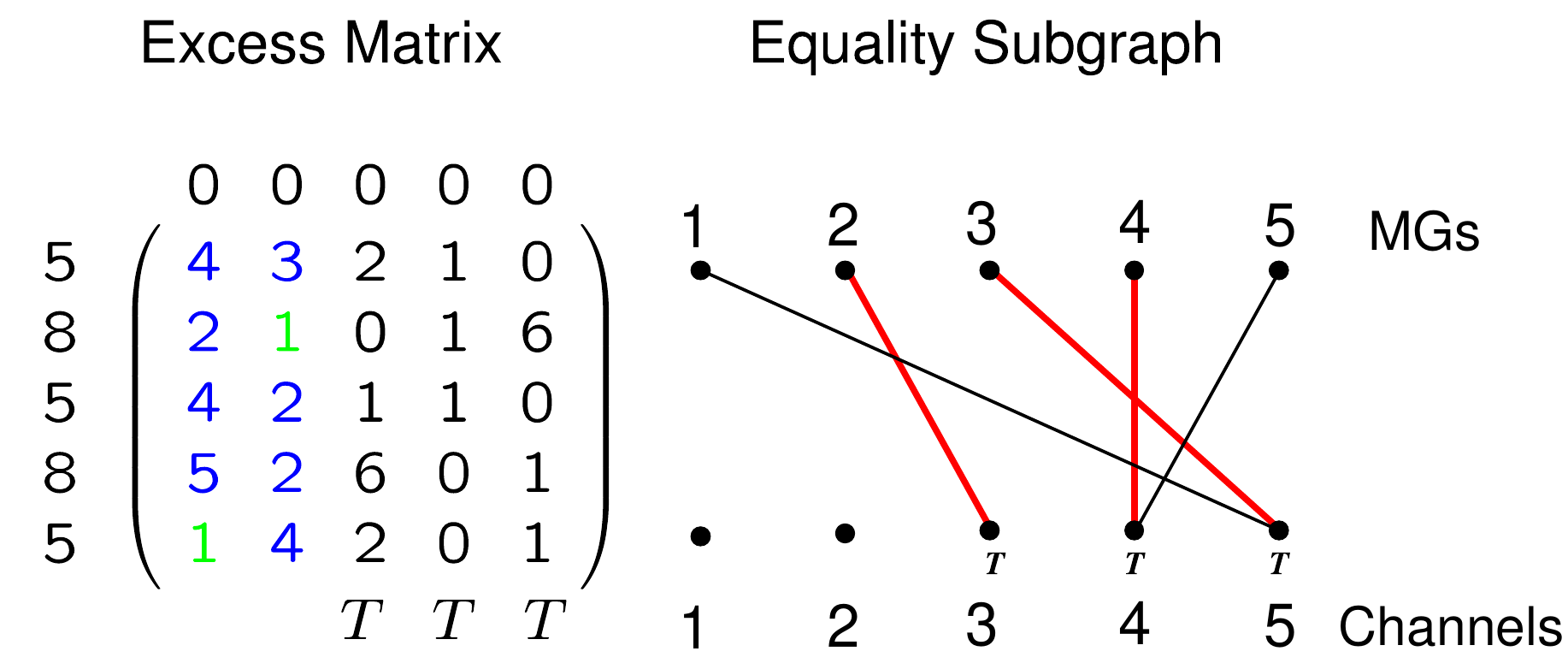}
\caption{The excess matrix and corresponding equality subgraph.}
\label{fig:excess_matrix_2}
\end{figure}

The initial values of $a_i = [5,8,5,8,5],$ (maximum of every row) $b_i = [0,0,0,0,0]$.  Initially, Q is first two columns, R = 0, and T = Y - Q. Therefore, T is last three columns of the $R_{k,g}$.
For excess matrix, the value of $e_{i,j}$ is calculated as $e_{i,j} = a_i + b_i - R_{i,j}$. Such as, for i=1,j=1, the value of $e_{1,1} = 5+0-1 = 4$. Therefore, the excess matrix and corresponding equality subgraph is illustrated in Figure \ref{fig:excess_matrix_2}.

\begin{figure}
\centering
\includegraphics[width=3.5in]{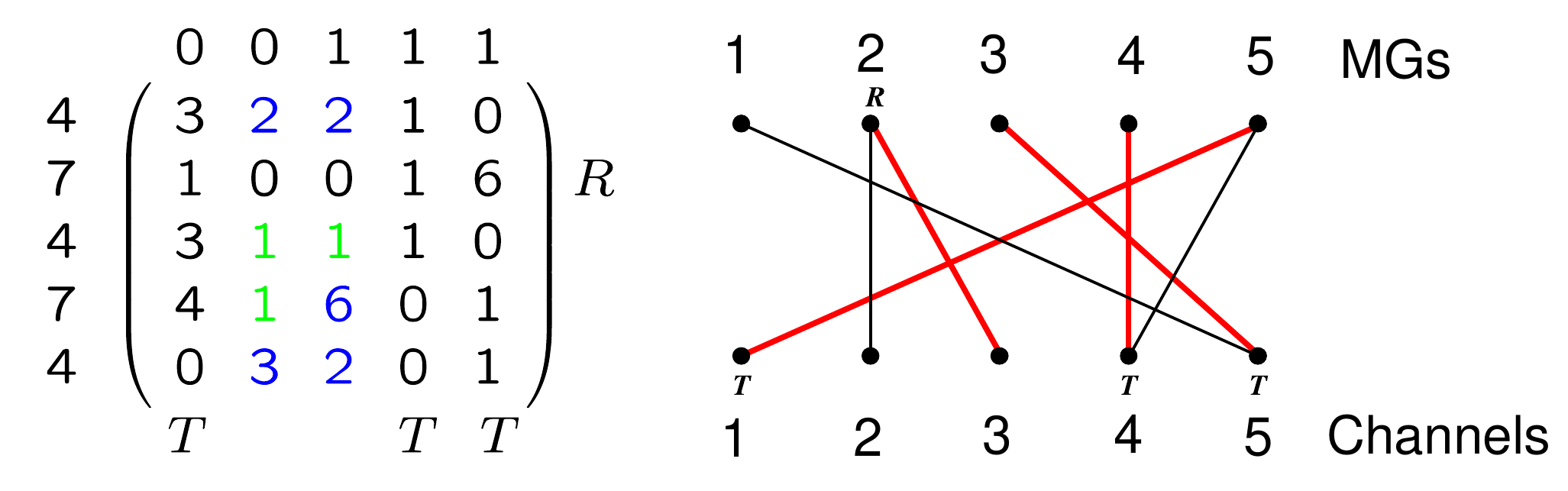}
\caption{The excess matrix and corresponding equality subgraph with $\delta =1$.}
\label{fig:excess_matrix_3}
\end{figure}

It can be observed that vertices 1 and 2 are not covered. Therefore, this is not an optimal solution. Now, value of $\delta$ is calculated as $\delta  = \min\left(a_i  + b_j - e_{i,j}\right),(5 + 0 - 4 = 1)$, and $[a_i] = [a_i] - \delta$ and $[b_i] = [b_i] + \delta$ are updated. $a_i$ is updated except for R matrix, and $b_i$ is updated for T matrix. Therefore, $a_i = [4,7,4,7,4], b_i = [0,0,1,1,1]$, R is second row, and T is first, fourth and fifth column. As MG 1 and MG 3 matches to a single channel. Therefore, it is not an optimal matching, as illustrating in Figure \ref{fig:excess_matrix_3}.

\begin{figure}
\centering
\includegraphics[width=3.5in]{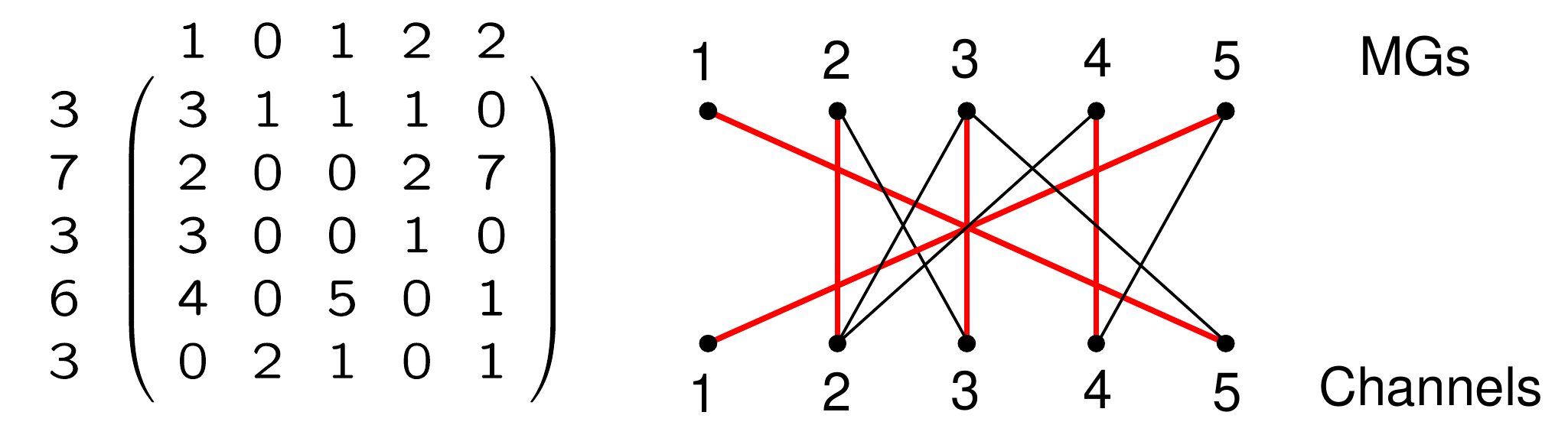}
\caption{The excess matrix and corresponding equality subgraph with $\delta =1$, and updated $a_i$, $b_j$.}
\label{fig:excess_matrix_4}
\end{figure}

 Again with $\delta = 1$, update $a_i = [3,7,3,6,3], b_i = [1,0,1,2,2]$, as illustrated in Figure \ref{fig:excess_matrix_4}. 

\begin{figure}
\centering
\includegraphics[width=3.5in]{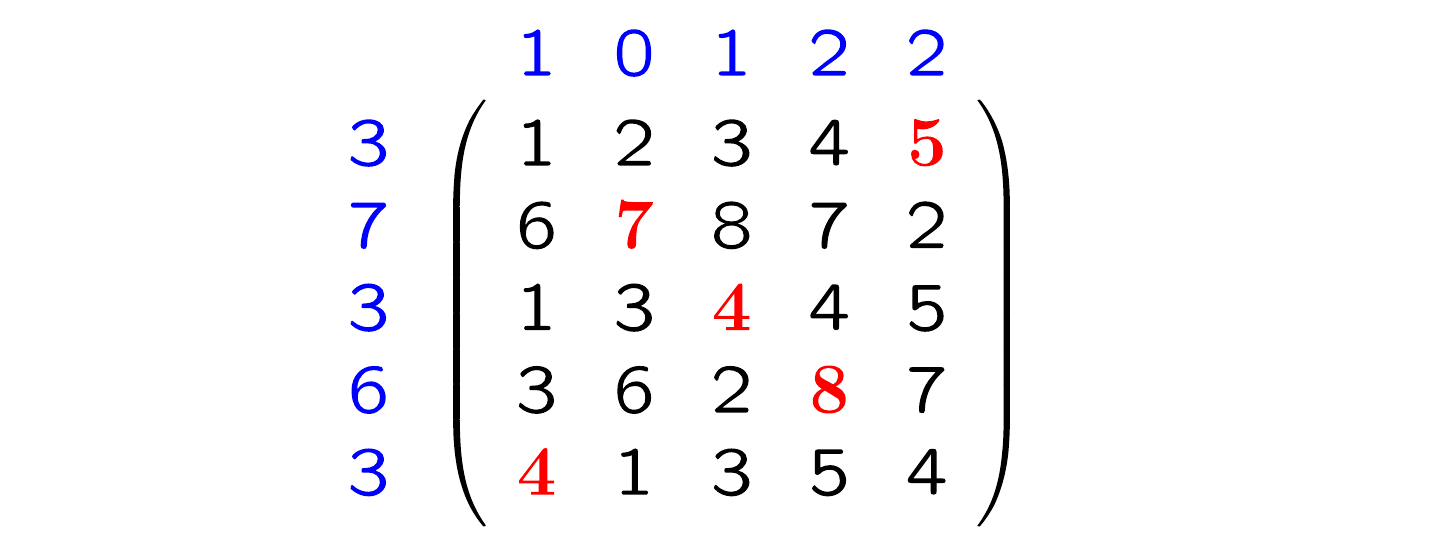}
\caption{The original weighted matrix with optimal matching.}
\label{fig:excess_matrix_5}
\end{figure}

As all vertices are covered with minimal number of edges, and all MGs find their respective optimal matches, this is an optimal matchin, with the optimal rate $28$ that can be obtained as (5+7+4+8+4 = 28), as illustrated in Figure~\ref{fig:excess_matrix_5}. 

\begin{figure}
\centering
\includegraphics[width=0.7\hsize]{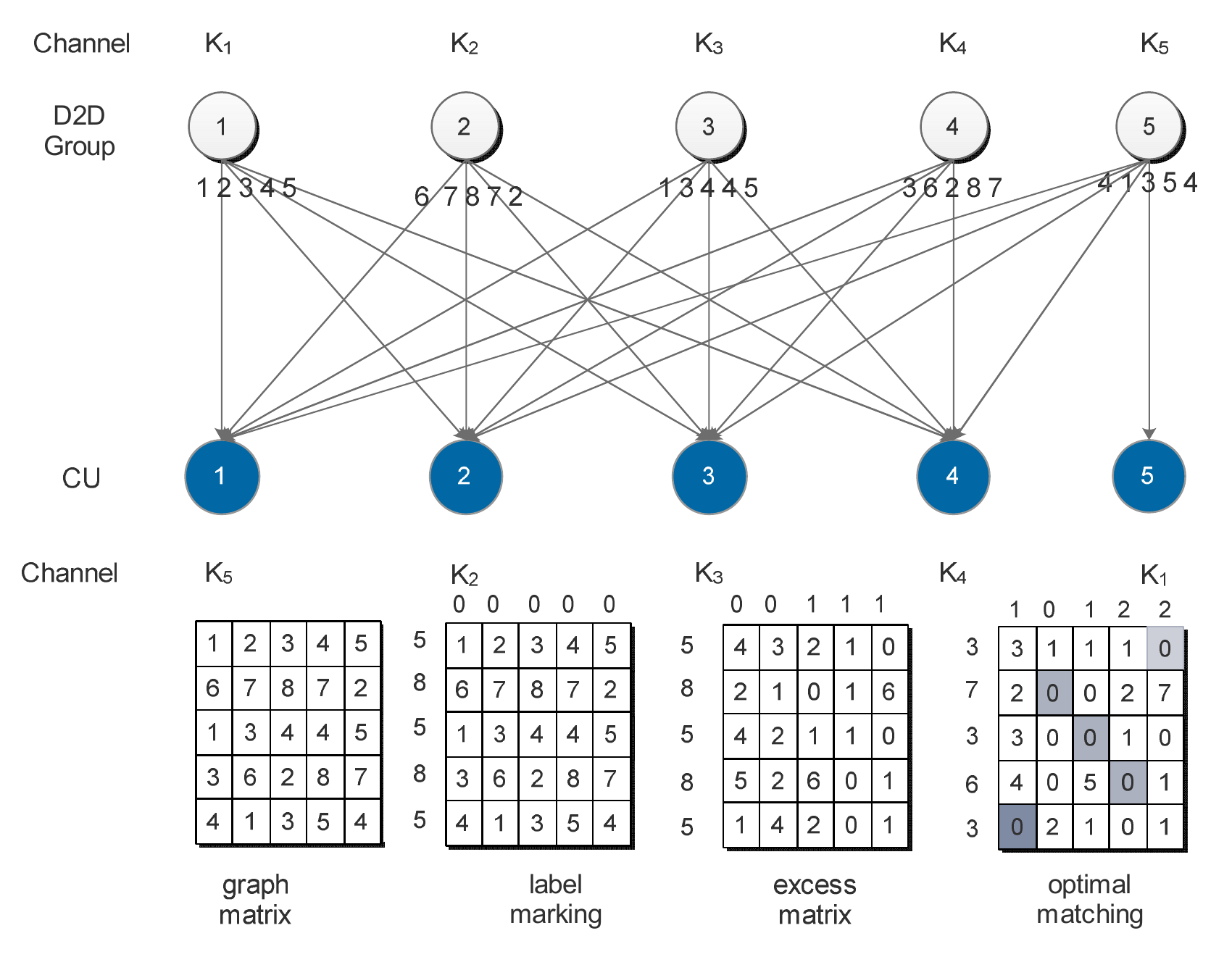}
\caption{Example for Bipartite Graph matching algorithm with G = 5 and C = 5.}
\label{fig:excess_matrix_6}
\end{figure}

A pictorial representation of outcome of different steps of Algorithm ~\ref{algo_bipartite} are given in Figure~\ref{fig:excess_matrix_6}.

\section{Proof of Lemma~\ref{ch_2_lemma1_MG_2_1}}
\label{appendices:lemma1}
\setcounter{equation}{0}
Let $\mathcal{G}_k$ be set of MGs sharing the resources with the $k^{\textrm{th}}$ CU. Let variables $p_{c,k}$ and $p_{g,k}, g \in \mathcal{G}_k$ denote the transmit power by CU and MG. Let $\overline{\mathbf{p_{opti}}}$ denotes the combine vector, $\overline{\mathbf{p_{opti}}} = \lbrace p_{c,k}, p_{g_1}^k,\ldots, p_{g_k}^k\rbrace$. The sum throughput maximization problem of $\mathcal{G}_{k}$ MGs can be expressed as follows:
\begin{align}
&\mathop{\max}_{\overline{\mathbf{p_{opti}}}} \sum_{g=1}^{\mathcal{G}_k} R_{g,r}^D \nonumber\\
\text{such that \ } &C_1 : \sum_{g\in \mathcal{G}_k} p_{g,k} h_{g,b}^k \leq I_\textrm{th}\left(p_{c,k}\right) \nonumber\\
&C_2: 0 \leq \sum_{g=1}^{\vert \mathcal{G}_k \vert} p_{g,k} \leq P_{g}^{max},\nonumber\\
& C_3: 0 \leq p_{c,k} \leq P_c^{\max}  
\end{align}
Let $\mathcal{W} = C_1 \cap C_2 \cap C_3$. As $C_1$, $C_2$ and $C_3$ are finite and closed region, therefore $\mathcal{W}$ is also finite closed domain and parameter $\overline{\mathbf{p_{opti}}}$ is also set accordingly to ensure that $\mathcal{W}$ is a non-empty set, therefore, we assume that $\mathcal{W}$ is closed and bounded set. Let $\partial \mathcal{W}$ be the boundary region of $\mathcal{W}$, therefore, $\mathcal{W}'$ = $\lbrace \mathcal{W}\rbrace \backslash \lbrace \partial \mathcal{W}\rbrace$. Further consider two power values, one is inside the $\mathcal{W}'$ i.e $\overline{\mathbf{p_{opti}}^*}$, and another point on boundary $\overline{\mathbf{p_{opti}}^b}$. Let $\overline{\mathbf{p_{opti}}^b} = \alpha \overline{\mathbf{p_{opti}}^*}$ and $\alpha > 1$. The throughput expression after replacing $\overline{\mathbf{p_{opti}}^*}$ with $ \alpha\overline{\mathbf{p_{opti}}^*}$ can be written as 
\begin{align*}
&C\left(\alpha\overline{\mathbf{p_{opti}}^*} \right) = {\log_2} \left( \left(1+ \beta_{2} \right) \right) > C\left(\overline{\mathbf{p_{opti}}^*}\right),
\end{align*}
\begin{align*}
\textrm{where}, \beta_{2} =  \dfrac{\overline{\mathbf{p_{opti}}^*} h_{g,r}^k} {\dfrac{N_0}{\alpha} + p_{c,k} h_{c,r}^k + \sum_{j \in \mathcal{G}_k, j\neq g} h_{j,g}^{k}} 
\end{align*}
With increasing $\alpha$, $C\left(\alpha\overline{\mathbf{p_{opti}}^*} \right) > C\left(\overline{\mathbf{p_{opti}}^*} \right)$ which contradict the assumption that $\left(\overline{\mathbf{p_{opti}}^*} \right)$ is optimal solution, thus $\left(\overline{\mathbf{p_{opti}}^*} \right) \in \partial{\mathcal{W}}$ holds. 
Therefore either of D2D-Txs $\mathcal{G}_k$ or the $k^\textrm{th}$ CU  will transmit at maximum power for maximizing system throughput.

\section{Proof of Lemma~\ref{ch_2_lemma1_MG_2_2}}
\label{ch2_appendices_2}
\setcounter{equation}{0}
A convex function $f$ is quasi-convex \textit{iff} 
\begin{align}
f\left(p'_1,p'_2,\ldots, p'_n\right)\leq f\left(p_1, p_2, \ldots, p_n\right)
\Rightarrow &\triangledown f\left(p_1, p_2, \ldots, p_n\right)^T \nonumber\\&\left( p'_1 - p_1, p'_2 - p_2,\ldots, p'_n - p_n\right) \leq 0  \label{eq:append_2_1}
\end{align}
Equation \eqref{main_problem} shows that $\mathcal{P}_1$ is the sum of $K$
individual functions.  As we know that addition and differentiation properties do not change the SINR bounds, we prove that SINR of every CU or D2D user is quasi-convex, then their sum is also quasi-convex. The sum SINR expression can be written as 
\begin{align}
\textrm{SINR} =  \sum_{i=1}^{C} \dfrac{p_i}{\sum_{j \neq i} p_j +  N_0}\label{eq:append_2_e1}
\end{align}
As we are considering an interference limited system, therefore, we may omit the constant $N_0$ in \eqref{eq:append_2_e1}. In addition, it does not impact the convexity or shape of the individual function. 
Now, we have two cases: 1) either power in numerator is constant or 2) it is varying. 
\begin{align}
\textrm{SINR} = \dfrac{\textrm{Constant}}{p_1+ p_2 + \ldots + p_{C}}\label{eq:append_2_2}
\end{align}
The shape of \eqref{eq:append_2_2} for varying power 
$p_1, \ldots p_{C}$ is hyperbolic, and therefore, quasi-convex. Now the second case - numerator ($p_1$) is variable
  \begin{align*}
 \textrm{SINR} = \dfrac{p_1}{p_2 + \ldots + p_{C - 1}}  \label{eq:append_2_3}
\end{align*}
From second inequality in \eqref{eq:append_2_1},
\begin{align}
&\triangledown \textrm{SINR} \left(p_1, \ldots ,p_{C}\right)^T \left(p'_1 - p_1, \ldots, p^{'}_{C} -  p_{C}\right)^T \nonumber\\
&=\left(\frac{1}{p_2 + \ldots + p_{C}}, \ldots , \dfrac{-p_1}{\left(p_2+ \ldots + p_{C}\right)^2}	\right)\times \left(p_1^{'} - p_1, \ldots, p_{C}^{'} - p_{C} \right)^T \nonumber\\
\end{align}
\begin{align}
& = \dfrac{p_1^{'} -  p_1}{p_2 + \ldots + p_{C}} -  p_1 \sum_{i=2}^{C} \dfrac{p_i^{'} -  p_i}{\left(p_2 + \ldots + p_{C}\right)^2} \leq 0, \nonumber\\
&= p_1^{'} \left(p_2 + \ldots + p_{C}\right) \leq p_1 \left(p_2^{'} + \ldots + p_{C}^{'}\right), \Rightarrow \dfrac{p_1^{'}}{p_2^{'}+ \ldots + p_{C}^{'}} \leq \dfrac{p_1}{p_2+ \ldots + p_{C}}    \label{eq:append_2_4}
\end{align}
Equation \eqref{eq:append_2_4} is the first inequality in \eqref{eq:append_2_1}. This shows for any combination of power, 
the sum SINR is quasi-convex, and therefore, the maximum value lies only on the corner points. 

\section{Example for corner search method for the power allocation for the scenario when $G_k$ = 2}
\label{appndx:examples_2}
\setcounter{equation}{0}
Let consider a case where two MGs share the channel with a CU. Let the maximum value of transmission power is $P_c^{\max} =  P_{g_1}^{\max} =P_{g_2}^{\max} = 1$ watt, $\gamma_r^{\textrm{th}}$ = $\gamma_{r'}^{\textrm{th}}$  = $\gamma_c^{\textrm{th}}$ =3 watt or 5 dB.
For region 1, let $p_{c,k}$ =  $P_c^{\max}$ = 1 watt. Using these parameter values, equations  \eqref{subsection_eq_5}-\eqref{subsection_eq_7} reduce to 
\begin{subequations}
\begin{align}
(2.9*e^{-5}) p_{g_1} + (-3*3e^{-7})  p_{g_2} &= 3* 4.5e^{-6} \label{bi_example_1}\\
(-3*3.2e^{-7})p_{g_1} + (2.9e^{-5}) p_{g_2} &= 3* 5e^{-6} \label{bi_example_2}\\
(-3*2.8e^{-7})p_{g_1}  + (-3*3.22e^{-6}) p_{g_2} &= - 5.2e^{-6}\label{bi_example_3};
\end{align}
\end{subequations}
By solving~\eqref{bi_example_1}-\eqref{bi_example_3}, we get three values of $p_{g_1,k}$ and 
$p_{g_2,k}$, $\{$ $[0.4809,0.4965]$, $[0.1754,0.5230]$, $[0.4821,0.5332]$ $\}$

Similarly, for region 2, let $p_{g_1}$ = $P_{g_1}^{\max}$ = 1 watt, equations  \eqref{subsection_eq_5}-\eqref{subsection_eq_7} reduce to  
\begin{subequations}
\begin{align}
 - (3*3e^{-7}) * p_{g_2} - 3*4.5e^{-6} * p_{c} &= - 2.9e^{-5} \label{bi_example_4}\\
 (2.9e^{-5}) * p_{g_2} - 3*3.2e^{-6}* p_{c}  &= 3*3.2e^{-7}\label{bi_example_5} \\
  - (3*3.22e^{-6}) * p_{g_2}  - 5.2e^{-6} p_{c} &= 3*2.8e^{-7} \label{bi_example_6};
\end{align}
\end{subequations}

By solving~\eqref{bi_example_4}-\eqref{bi_example_6} , we get three values of $p_{c}$,  $p_{g_2}$ = $\{[0.316, 0.044]$, $[0.732, 0.021]$, $[0.037, 0.25]\}$

For region 3, let $p_{g_2}$ = $P_{g_2}^{\max}$ = 1 watt, equations  \eqref{subsection_eq_5}-\eqref{subsection_eq_7} reduce to 
\begin{subequations}
\begin{align}
 - (3*4e^{-7}) * p_{g_1,k} - 3*3.56e^{-6} * p_{c,k} &= - 3.43e^{-5} 
\label{bi_example_7} \\
 (5.2e^{-5}) * p_{g_1,k} - 3*2.7e^{-6}* p_{c,k}  &= 3*2.5e^{-7} \label{bi_example_8}\\
  - (3*2.76e^{-6}) * p_{g_1,k}  - 4.78^{-6} p_{c,k} &= 3*3.42e^{-7}\label{bi_example_9}
\end{align}
\end{subequations}

By solving~\eqref{bi_example_7}-\eqref{bi_example_9}, we get three values of $p_{c,k}$,  $p_{g_1,k}$ = $\{$ $[0.678, 0.022]$, $[0.6805, 0.5103]$, $[0.1435, 0.703]$ $\}$

For region 4, let $p_{c}$ = $P_c^{\max}$, and $p_{g_1}$  = $P_g^{\max}$, then solving equation \eqref{subsection_eq_6} for $p_{g_2}$,
$p_{g_2}$ = $((3*2.9*e^{-7}) + (3*4.2*e^{-6})) / (4.2*e^{-5}) = 0.3207$watt.

For region 5, let $p_{c}$ = $P_c^{\max}$, and $p_{g2}$  = $P_g^{\max}$,  then solving equation \eqref{subsection_eq_5} for $p_{g_1}$,
$P_{g1}$ = $((3*5.3*e^{-7}) + (3*3.5*e^{-7})) / (2.5*e^{-5}) = 0.1056$watt.

For region 6, let $p_{g1}$ = $P_{g1}^{\max}$, and $p_{g2}$  = $P_{g2}^{\max}$ then solving equation \eqref{subsection_eq_7} for $p_{c}$,
$p_{c}$ = $((3*4.7*e^{-7}) + (3*5.6*e^{-6})) / (3.8*e^{-5}) = 0.4737$watt.

Region 7, where both MGs and CU transmits at the maximum power, that is, $p_{g1}$ = $P_{g1}^{\max}$, and $p_{g_2}$  = $P_{g_2}^{\max}$ ,  $p_{c}$ = $P_c^{\max}$ = 1, SINR constraints of the MGs are not fulfilled. 
Therefore, to get an approximate optimal solution, we need to only trace these points, $\Big( p_c, p_{g1}, p_{g2}\Big)$, $\Big([1,0.4809,0.4965]$, $[1,0.1754,0.5230]$, $[1,0.4821,0.5332]$ $[0.3157, 1,0.0436]$, $[0.732, 1, 0.021]$, $[0.037, 1, 0.25]$ $[0.678, 0.022, 1]$,

$[0.6805, 0.5103, 1]$, $[0.1435, 0.703, 1]$ $[1, 1, 0.3207]$, $[1, 0.1056, 1]$,
$[0.4737, 1, 1]\Big)$,

\bibliographystyle{IEEEtran}
\linespread{1}

\end{document}